\documentclass[aps,prb,amsmath,amssymb,reprint,superscriptaddress,preprintnumbers,showpacs,intlimits]{revtex4-1}
\pdfoutput=1
\usepackage{bm,latexsym,mathrsfs,enumerate,color}
\usepackage[mathcal]{euscript}
\usepackage{hyperref}
\usepackage{graphicx}
\usepackage[normalem]{ulem}

\renewcommand{\vec}[1]{\bm{#1}}
\begin{document}

\title{Magnetic vortex-antivortex crystals generated by spin-polarized current}

\author{Yuri Gaididei}
 \affiliation{Bogolyubov Institute for Theoretical Physics, 03143 Kiev, Ukraine}

\author{Oleksii M. Volkov}
\affiliation{Taras Shevchenko National University of Kiev, 01601 Kiev, Ukraine}

\author{Volodymyr P. Kravchuk}
 \affiliation{Bogolyubov Institute for Theoretical Physics, 03143 Kiev, Ukraine}

\author{Denis D. Sheka}
     \email{sheka@univ.net.ua}
\affiliation{Taras Shevchenko National University of Kiev, 01601 Kiev, Ukraine}

\date{\today}

%
%

\begin{abstract}
 We study vortex pattern formation in thin ferromagnetic films under the action of strong spin-polarized currents. Considering the currents which are polarized along the normal of the film plane, we determine  the critical current above which the film goes to a saturated state with all magnetic moments being perpendicular to the film plane. We show that  stable square vortex-antivortex superlattices
 (\emph{vortex crystals}) appears slightly below the critical current.  The melting of the vortex
 crystal occurs with  current further decreasing. A mechanism of current-induced periodic vortex-antivortex lattice formation  is proposed. Micromagnetic simulations confirm our analytical results with a high accuracy.
\end{abstract}




\maketitle


\section{Introduction}
\label{sec:intro}

The spin-polarized current is a convenient means to handle magnetization states of nanomagnets without applying of external magnetic field\cite{Lindner10}. That is of high applied importance for constructing purely current controlled devices\cite{Bohlens08,Drews09}. One of the effective way to influence the film magnetization by the spin-polarized current is to use the pillar structure, where the current flows perpendicular to the magnetic film.\cite{Myers99,Katine00,Myers02,Kiselev03,Li04,Maicas04} Special efforts in this way were made to explore the possibility to control the properties of magnetic vortex\cite{Caputo07,Ivanov07b,Sheka07b,Pribiag07,Choi07a,Choi08,Jin09,Khvalkovskiy09a,Choi10,Jaromirska11,Gaididei10} because the latter is a convenient carrier of bit of information. The theoretical study in this way is based on the Slonczewski-Berger model.\cite{Slonczewski96,Berger96,Slonczewski02}

In this paper we focus on the problem of regular pattern formation (vortex-antivortex superlattice) under the action of strong spin-polarized currents, which precedes the saturation. Superstrucures of vortices are known from ages of Kelvin's fluid vortices. \cite{Thomson10} Nowadays superlattices of vortices are known in superconductivity \cite{Abrikosov04}, superfluidity \cite{Donnelly91}, Bose--Einstein condensates (rotating \cite{Fetter09}, nonrotating \cite{Ruben10}, optically dressing condensate\cite{Komineas12}), and optics \cite{Dreischuh02,Eastwood12,Xavier12}.  Vortex--like superstructures appear also in magnetism: skyrmion crystals were predicted in chiral magnets \cite{Roessler06}, which is now well--confirmed experimentally \cite{Muehlbauer09,Yu10,Yu11}, vortex-antivortex lattice (chirality waves) appears in Kondo lattice model\cite{Solenov12}. Recently we found vortex--antivortex superlattices (\emph{vortex crystals}) in nanomagnets under the action of strong spin-polarized current.\cite{Volkov11} Using micromagnetic simulations we found that crystallization precedes a saturation: the square superlattices were observed for a range of current densities in immediate vicinity of $J_c$, which is the critical current which saturates the magnetization along its direction.\cite{Volkov11} Here we prove theoretically the possibility of vortex-antivortex superlattices in ordinary isotropic magnetic film. To this end, we build the full theory of saturation of a thin ferromagnetic film by transverse spin-polarized current. In particular, we show that loss of stability of the saturated state leads to appearance of the stable square vortex crystals.

The paper is organised as follows: In Sec.~\ref{sec:model} we describe the model and our approach.
The linear analysis (Sec.~\ref{sec:Harmonic}) enables us to obtain the value of the saturation current $J_c$ as function of material parameters and the film thickness. The nonlinear analysis (Sec.~\ref{sec:nonlin}) proves the possibility of stable square vortex-antivortex superlattices in pre-saturated regime. All the obtained analytical results we check with micromagnetic simulations (Sec.~\ref{sec:simuls}). Besides, using the simulations we describe the transition from crystal phase into fluid phase which appears with the current decrease.

\section{Model and discrete description}
\label{sec:model}

We consider here a soft magnetic film with thickness $h$ and lateral size $L\gg h$. Magnetization of the film we model as a three-dimensional cubic lattice of magnetic moments $\vec M_{\vec \nu}$ with lattice spacing $a\ll h$, where $\vec\nu=a(\nu_x,\nu_y,\nu_z)$ with $\nu_x,\nu_y,\nu_z\in\mathbb{Z}$ is a three-dimensional index\footnote{Everywhere in the text we denote the two dimensional indexes by Latin letters while the three dimensional indexes are denoted by Greek letters.}. In the following we use the notations $\mathcal{N}_z=h/a$ and $\mathcal{N}_{xy}=L^2/a^2$ for number of magnetic moments along thickness and within the film plane respectively.  We assume also that the magnetization of the film is uniform along thickness. That enable us to base our study on the two-dimensional discrete Landau-Lifshitz-Slonczewski equation:\cite{Slonczewski96,Berger96,Slonczewski02}
\begin{equation} \label{eq:LLS}
\dot{\vec{m}}_{\vec n} = \vec m_{\vec n}\times{\partial\mathcal{E}}/{\partial\vec{m}_{\vec n}}-j\varepsilon \vec m_{\vec n}\times[\vec m_{\vec n}\times\hat{\vec z}],
\end{equation}
which describes the magnetization dynamics under influence of spin-polarized current which flows perpendicularly to the magnet plane, along $\hat{\vec z}$-axis. It is also assumed that the current flow and its spin-polarization are of the same direction in \eqref{eq:LLS}. The two dimensional index $\vec n=a(n_x,\,n_y)$ with $n_x,n_y\in\mathbb{Z}$ numerates the normalized magnetic moments $\vec m_{\vec n} = \vec M_{\vec n}/|\vec M_{\vec n}|$ within the film plane. The overdot indicates derivative with respect to the rescaled time in units of $(4\pi\gamma M_s)^{-1}$, $\gamma$ is gyromagnetic ratio, $M_s$ is the saturation magnetization, and $\mathcal{E}=E/(4\pi M_s^2 a^3 \mathcal{N}_z)$ is dimensionless magnetic energy. The normalized electrical current density $j=J/J_0$, where $J_0=M_s^2|e|h/\hbar$ with $e$ being electron charge and $\hbar$ being Planck constant.  The spin-transfer torque efficiency function $\varepsilon$ has the form $\varepsilon={\eta\Lambda^2}/{\left[(\Lambda^2+1)+(\Lambda^2-1)(\vec m\cdot\hat{\vec z})\right]}$, where $\eta$ is the degree of spin polarization and parameter $\Lambda\geqslant1$ describes the mismatch between spacer and ferromagnet resistance \cite{Slonczewski02,Sluka11}. To simplify representation we omitted damping in the equation of motion \eqref{eq:LLS}, since the role of damping is not essential for crystallization of vortices; moreover, the saturation current does not depend on the damping constant.\cite{Volkov11}

The total energy of the system $E=E_\mathrm{ex}+E_\mathrm{d}$ consists of two parts: exchange and dipole-dipole contributions. The exchange energy has the form
\begin{equation} \label{eq:Eex}
E_\mathrm{ex}=-\mathcal{S}^2\mathcal{N}_z\sum\limits_{\vec n,\vec l\ne\vec0}\mathcal{J}_{\vec l}\vec m_{\vec n}\cdot\vec m_{\vec n+\vec l},
\end{equation}
where $\vec n$, $\vec l$ are two-dimensional indexes, $\mathcal{S}$ is value of spin of a ferromagnetic atom, and $\mathcal{J}_{\vec l}$ denotes the exchange integral between atoms distanced on $\vec l$.

The energy of dipole-dipole interaction is
\begin{equation} \label{eq:Ems}
\begin{split}
E_\mathrm{d}=\frac{M_s^2a^6}{2}&\sum\limits_{\vec\nu\ne\vec\lambda}\biggl[\frac{ (\vec m_{\vec\nu}\cdot \vec m_{\vec\lambda})}{|\vec\lambda-\vec\nu|^3}\\
&-3\frac{\left(\vec m_{\vec\nu}\cdot (\vec\lambda-\vec\nu)\right) \left(\vec m_{\vec\lambda} \cdot (\vec\lambda-\vec\nu)\right)}{|\vec\lambda-\vec\nu|^5}\biggr],
\end{split}
\end{equation}
where $\vec\lambda$ and $\vec\nu$ are three dimensional indexes.

By introducing the complex variable
\begin{equation} \label{eq:psi}
\psi_{\vec n}=\frac{m^x_{\vec n}+im^y_{\vec n}}{\sqrt{1+m^z_{\vec n}}},
\end{equation}
one can write the Eq. \eqref{eq:LLS} in form
\begin{equation} \label{eq:LLS-psi}
i\dot\psi_{\vec n}=-\frac{\partial\mathcal{E}}{\partial\psi_{\vec n}^*}-i\varkappa\frac{1-\frac12|\psi_{\vec n}|^2}{1-\frac\xi2|\psi_{\vec n}|^2}\psi_{\vec n},
\end{equation}
where $\varkappa=j\eta/2$ is renormalized current, $\xi=1-\Lambda^{-2}$ and $\psi^*$ denotes the complex conjugation of $\psi$.

It is well known that in the absence of driving ($\varkappa=0$) the spatially homogeneous state with all moments lying in the $xy$-plane (easy-plane magnetic state) is the most
energetically favorable state of a thin ferromagnetic film. On the other hand, as it is seen from Eqs.~\eqref{eq:Eex}, \eqref{eq:Ems} and \eqref{eq:LLS-psi} for large positive $\varkappa$ the stationary state of the system
corresponds to $\psi_{\vec n} = 0$ or in other words, the system goes to the state when all magnetic moments are oriented along the $z$-axis (saturated state). This means that there
should exist a critical current $\varkappa_c$ below which the saturated state loses its stability. Our goal is to study the
behavior of the system near threshold of stability of the saturated state. Near the threshold $m^z_{\vec n}\lesssim 1$ and $|\psi_{\vec n}|\ll1$, hence one can expand components of the magnetization vector into series in the way similar to the representation in terms of the Bose operators:\cite{Akhiezer68}
\begin{equation} \label{eq:mxmymz}
\begin{split}
&m^x_{\vec n}=\frac{\psi_{\vec n}+\psi_{\vec n}^*}{\sqrt{2}}\left(1-\frac{|\psi_{\vec n}|^2}{4}\right)+\mathcal{O}(|\psi_{\vec n}|^5)\\
&m^y_{\vec n}=\frac{\psi_{\vec n}-\psi_{\vec n}^*}{i\sqrt{2}}\left(1-\frac{|\psi_{\vec n}|^2}{4}\right)+\mathcal{O}(|\psi_{\vec n}|^5)\\
&m^z_{\vec n}=1-|\psi_{\vec n}|^2.
\end{split}
\end{equation}

Substituting \eqref{eq:mxmymz} into \eqref{eq:LLS-psi} one obtains the equation of motion accurate to terms of the third order
\begin{equation} \label{eq:LLS-psi-nonlin}
i\dot\psi_{\vec n}=-\frac{\partial\mathcal{E}}{\delta\psi_{\vec n}^*}-i\varkappa\psi_{\vec n}\left(1-\frac{1}{2\Lambda^2}|\psi_{\vec n}|^2\right).
\end{equation}

For the future analysis it is convenient to proceed to the wave-vector representation using the two-dimensional discrete Fourier transform
\begin{subequations}\label{eq:Fourier-def}
\begin{align}
\label{eq:four-inv}&\psi_{\vec n}=\frac{1}{\sqrt{\mathcal{N}_{xy}}}\sum\limits_{{\vec k}}\hat\psi_{\vec k}e^{i \vec k\cdot\vec n},\\
\label{eq:four}&\hat\psi_{\vec k}=\frac{1}{\sqrt{\mathcal{N}_{xy}}}\sum\limits_{{\vec n}}\psi_{\vec n}e^{-i \vec k\cdot\vec n}
\end{align}
\end{subequations}
with the orthogonality condition
\begin{equation} \label{eq:orth-cond}
\sum\limits_{{\vec n}}e^{i(\vec k-\vec k')\cdot\vec n}=\mathcal{N}_{xy}\Delta(\vec k-\vec k'),
\end{equation}
where $\vec k=(k_x,k_y)\equiv\frac{2\pi}{L}(l_x, l_y)$ is two-dimensional discrete wave vector, $l_x, l_y\in\mathbb{Z}$, and $\Delta(\vec k)$ is the Kronecker delta.
Applying \eqref{eq:Fourier-def} to the equation \eqref{eq:LLS-psi-nonlin} one obtains equation of motion in reciprocal space:
\begin{equation} \label{eq:main-Four}
\begin{split}
&-i\dot{\hat{\psi}}_{\vec k}=\frac{\partial\mathcal{E}}{\partial\hat\psi_{\vec k}^*}+i\frac{\partial\mathcal{F}}{\partial\hat\psi_{\vec k}^*},
\end{split}
\end{equation}
where the dimensionless energy of the system can be represented as a sum
\begin{equation} \label{eq:all-energies}
\mathcal{E}=\underbrace{\mathcal{E}^0_\mathrm{ex}+\mathcal{E}^0_\mathrm{d}}_{\mathcal{E}^0}+\underbrace{\mathcal{E}^\mathrm{nl}_\mathrm{ex}+\mathcal{E}^\mathrm{nl}_\mathrm{d}}_{\mathcal{E}^\mathrm{nl}}
\end{equation}
Here the term $\mathcal{E}^0=\mathcal{E}^0_\mathrm{ex}+\mathcal{E}^0_\mathrm{d}$ is the harmonic part of the energy. It consists of the exchange contribution
\begin{subequations}\label{eq:harmonic}
\begin{align}
\label{eq:Eex-lin}\mathcal{E}_\mathrm{ex}^0=&\ell^2\sum\limits_{\vec k}|\hat\psi_{\vec k}|^2k^2,
\end{align}
and the dipole-dipole contribution
\begin{equation} \label{eq:Ed-lin}
\begin{split}
\mathcal{E}_\mathrm{d}^0=\sum\limits_{\vec k}&\left[\frac{g(kh)}{2}-1\right]|\hat\psi_{\vec k}|^2\\
&+\frac{g(kh)}{4}\left[\frac{(k^x-ik^y)^2}{k^2}\hat\psi_{\vec k}\hat\psi_{-\vec k}+\text{c.c.}\right].
\end{split}
\end{equation}
\end{subequations}
The nonlinear part of the energy is described by the term $\mathcal{E}^\mathrm{nl}=\mathcal{E}^\mathrm{nl}_\mathrm{ex}+\mathcal{E}^\mathrm{nl}_\mathrm{d}$, which consists of nonlinear exchange contribution
\begin{subequations}\label{eq:nonlin}

\begin{equation} \label{eq:Eex-nl}
\begin{split}
\mathcal{E}_\mathrm{ex}^{\mathrm{{nl}}} &= \frac{\ell^2}{4\mathcal{N}_{xy}}\sum\limits_{\vec k_1\vec k_2\vec k_3\vec k_4}\Biggl[
\mathfrak{A}(\vec k_1,\vec k_2)\,\hat\psi_{\vec k_1}\hat\psi_{\vec k_2}^*\hat\psi_{\vec k_3}\hat\psi_{\vec k_4}^* \\
&\times \Delta(\vec k_1-\vec k_2+\vec k_3-\vec k_4)+\text{c.c.}\Biggr],
\end{split}
\end{equation}
and dipole-dipole one:
\begin{equation} \label{eq:Ed-nl}
\begin{split}
&\mathcal{E}_\mathrm{d}^\mathrm{nl}=-\frac{1}{4\mathcal{N}_{xy}}\sum\limits_{\vec k_1\vec k_2\vec k_3\vec k_4}\Biggl[
\mathfrak{B}(\vec k_1,\vec k_2)\hat\psi_{\vec k_1}\hat\psi_{\vec k_2}^*\hat\psi_{\vec k_3}\hat\psi_{\vec k_4}^*\\
&\times \Delta(\vec k_1-\vec k_2+\vec k_3-\vec k_4)+ \mathfrak{C}(\vec k_1)\hat\psi_{\vec k_1}\hat\psi_{\vec k_2}^*\hat\psi_{\vec k_3}\hat\psi_{\vec k_4}\\
&\times \Delta(\vec k_1-\vec k_2+\vec k_3+\vec k_4)+ \text{c.c.}\Biggr].
\end{split}
\end{equation}
\end{subequations}
The characteristic length
\begin{equation}
\ell=\sqrt{\frac{\mathcal{S}^2}{4\pi M_s^2a^3}\sum\limits_{\vec n}\vec n^2\mathcal{J}_{\vec n}},
\end{equation}
which appears in \eqref{eq:harmonic} and \eqref{eq:nonlin} is so called exchange length. We introduced also the following functions
\begin{subequations}\label{eq:functions}
\begin{align}\label{eq:A-function}
&\mathfrak{A}(\vec k_1,\vec k_2)\equiv k_1^2-2(\vec k_1\cdot\vec k_2),\\
&\mathfrak{B}(\vec k_1,\vec k_2)\equiv g(|\vec k_1-\vec k_2|h)+\frac{g(k_1h)}{2}-1,\\
&\mathfrak{C}(\vec k)\equiv g(kh)\frac{(k^x-ik^y)^2}{2k^2},\\
&g(x)\equiv\frac{x+e^{-x}-1}{x}.
\end{align}
\end{subequations}
Details of deriving of the Hamiltonian \eqref{eq:all-energies}-\eqref{eq:nonlin} in the wave-vector space are placed into the Appendix~\ref{app:Energies}.

The function $\mathcal{F}$ represents an action of the spin-polarized
current. It consists of two parts
\begin{subequations} \label{eq:F-function}
\begin{equation} \label{eq:F-function-1}
\mathcal{F}=\mathcal{F}^0+\mathcal{F}^\mathrm{nl},
\end{equation}
with the harmonic contribution
\begin{equation} \label{eq:F-function-2}
\mathcal{F}^0=\varkappa\sum\limits_{\vec k}\hat\psi_{\vec k}^*\hat\psi_{\vec k}
\end{equation}
and the nonlinear part
\begin{equation} \label{eq:F-function-3}
\begin{split}
\mathcal{F}^\mathrm{nl}&=-\frac{\varkappa}{4\Lambda^2\mathcal{N}_{xy}}\sum\limits_{\vec k_1,\vec k_2,\vec k_3,\vec k_4}\Bigl[\hat\psi_{\vec k_1}\hat\psi_{\vec k_2}\hat\psi_{\vec k_3}^*\hat\psi_{\vec k_4}^*\\
&\times\Delta(\vec k_1+\vec k_2-\vec k_3-\vec k_4) \Bigr].
\end{split}
\end{equation}
\end{subequations}

Note that we are interested in a large scale behavior of the system and restrict attention to long-wave excitations.
Therefore Eqs.~\eqref{eq:harmonic}, \eqref{eq:nonlin} and \eqref{eq:functions} are written in the limit $ka\ll1$.
%


\section{Harmonic approximation}
\label{sec:Harmonic}

First, we discuss solutions of Eqs.~\eqref{eq:main-Four} in the harmonic approximation, since they already capture many essential aspects of the problem. By neglecting all nonlinear terms
in \eqref{eq:main-Four}, equations for the complex amplitudes $\hat\psi_{\vec k}$ and $\hat\psi_{-\vec k}^*$ can be written in the form
\begin{equation} \label{eq:eq-of-motion}
\begin{split}
-i\dot{\hat{\psi}}_{\vec k}   &=  \left[k^2\ell^2-1+\frac{g(hk)}{2}+i\varkappa\right]\hat\psi_{\vec k}\\
&+\frac{g(hk)}{2}\frac{(k^x-ik^y)^2}{k^2}\hat{\psi}^*_{-\vec k},\\
i\dot{\hat{\psi}}_{-\vec k}^* &= \left[k^2\ell^2-1+\frac{g(hk)}{2}-i\varkappa\right]\hat\psi_{-\vec k}^*\\
&+\frac{g(hk)}{2}\frac{(k^x+ik^y)^2}{k^2}\hat{\psi}_{\vec k}.
\end{split}
\end{equation}
The solutions of Eq.~\eqref{eq:eq-of-motion} have the form form
\begin{equation} \label{eq:psi-sol}
\hat\psi_{\vec k}(t)=\Psi_+e^{z_+(\vec k)t},\qquad\hat\psi_{-\vec k}^*(t)=\Psi_-e^{z_-(\vec k) t},
\end{equation}
 where $\Psi_{\pm}(\vec k)$ are time independent amplitudes and  the rate constants $z_{\pm}(\vec k)$ are given by
\begin{equation} \label{eq:dispersion}
z_{\pm}(\vec k)=-\varkappa\pm\tilde\varkappa(k),
\end{equation}
where the rate function $\tilde\varkappa(k)$ is given by
\begin{equation} \label{eq:crit-current}
\tilde\varkappa(k)=\sqrt{(1-k^2\ell^2)\left(k^2\ell^2+g(hk)-1\right)},
\end{equation}
First of all it should be noted that since $\varkappa>0$ than accordingly to \eqref{eq:dispersion} the current plays role of an effective damping. That explains why we omitted weak natural damping
in Eq.~\eqref{eq:LLS}. That also explains the previous numerical results, where the saturation and magnetization dynamics under the high spin-current influence were independent on the damping coefficient\cite{Volkov11}.

\begin{figure}
\includegraphics[width=\columnwidth]{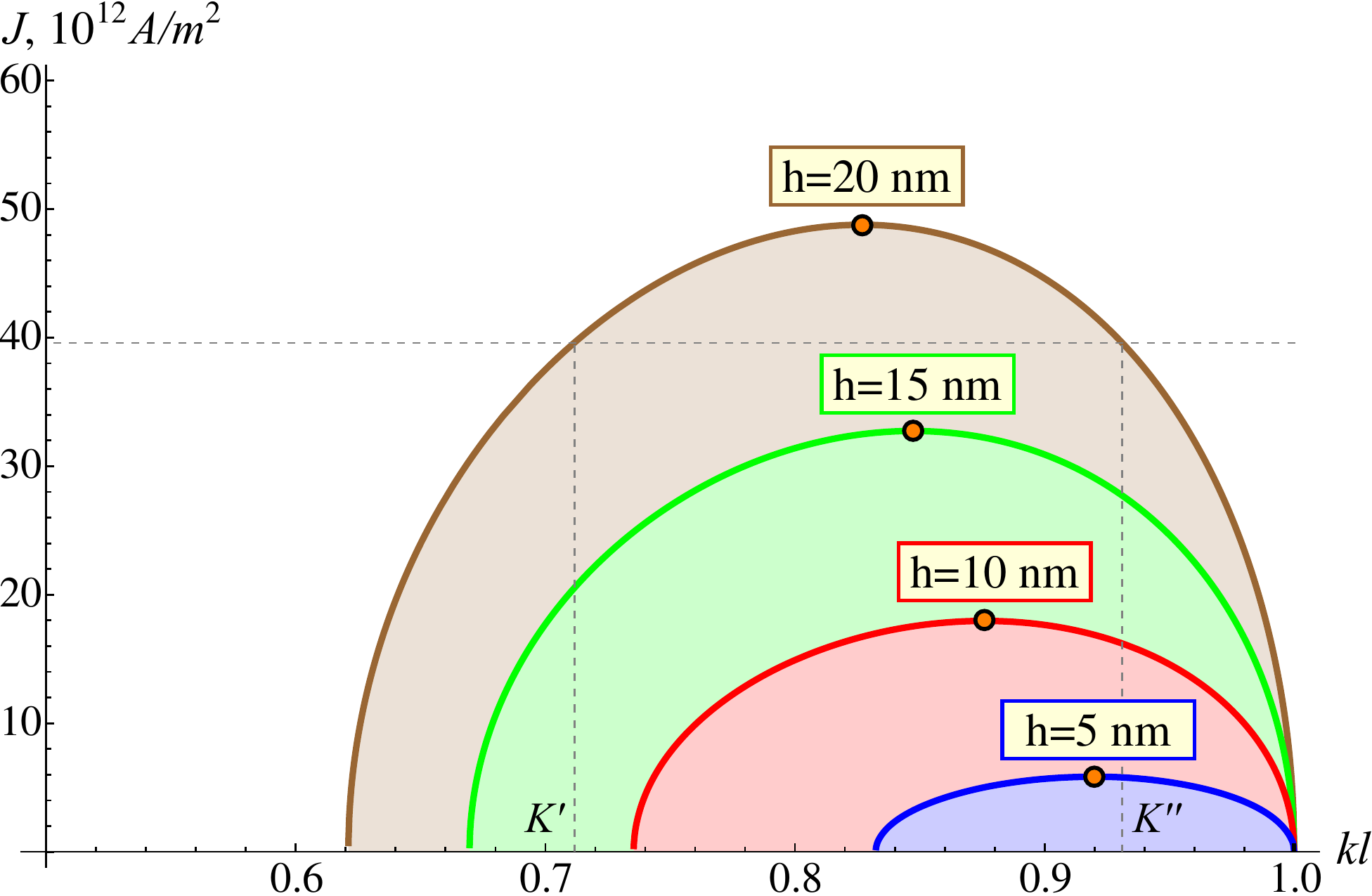}
\caption{Diagram of stability of the uniform state saturated transversally by spin current. The regions of instability are shown by filling and they are determined by condition $\varkappa<\tilde\varkappa$, where the the normalized current $\varkappa$ is rescaled to the real current density $J$. Thus for the given current value $J<J_c$ one has the range $[K',\,K'']$ of the instable wave-vectors.  Parameters of material and spin-current were taken the same as for simulations (see Section~\ref{sec:simuls}). Points show the maximums of dependencies $J(k\ell)$ and they determine the saturation current for the given thickness. }\label{fig:stab-diagramm}
\end{figure}

Function $\tilde{\varkappa}(k)$  is a non-monotonic one which reaches its maximum value $\varkappa_c$  at $k=K$:
\begin{equation}\label{eq:crit-current-max}
\frac{d \tilde\varkappa(K)}{d K}=0\;\qquad \varkappa_c=\max\limits_k\tilde\varkappa(k)\equiv\tilde\varkappa(K)\;.
\end{equation}
Typical shapes of the rate functions \eqref{eq:crit-current} are presented in Fig.~\ref{fig:stab-diagramm}.

For strong  currents when $\varkappa > \varkappa_c$, we have $\mathrm{Re}\,z_{\pm}(\vec k)<0$ for all values of the wave vector $\vec k$. This means that the stationary state of the system is the saturated state with $m_z=1$. However, for $\varkappa < \varkappa_c$ the  saturated state is linearly unstable with respect to modes $\hat{\psi}_{\vec k}$ with wave vectors close to the threshold wave vector $K$. The corresponding instability domains for different thicknesses are shown in the Fig.~\ref{fig:stab-diagramm} as filled regions.


\begin{figure}
\includegraphics[width=\columnwidth]{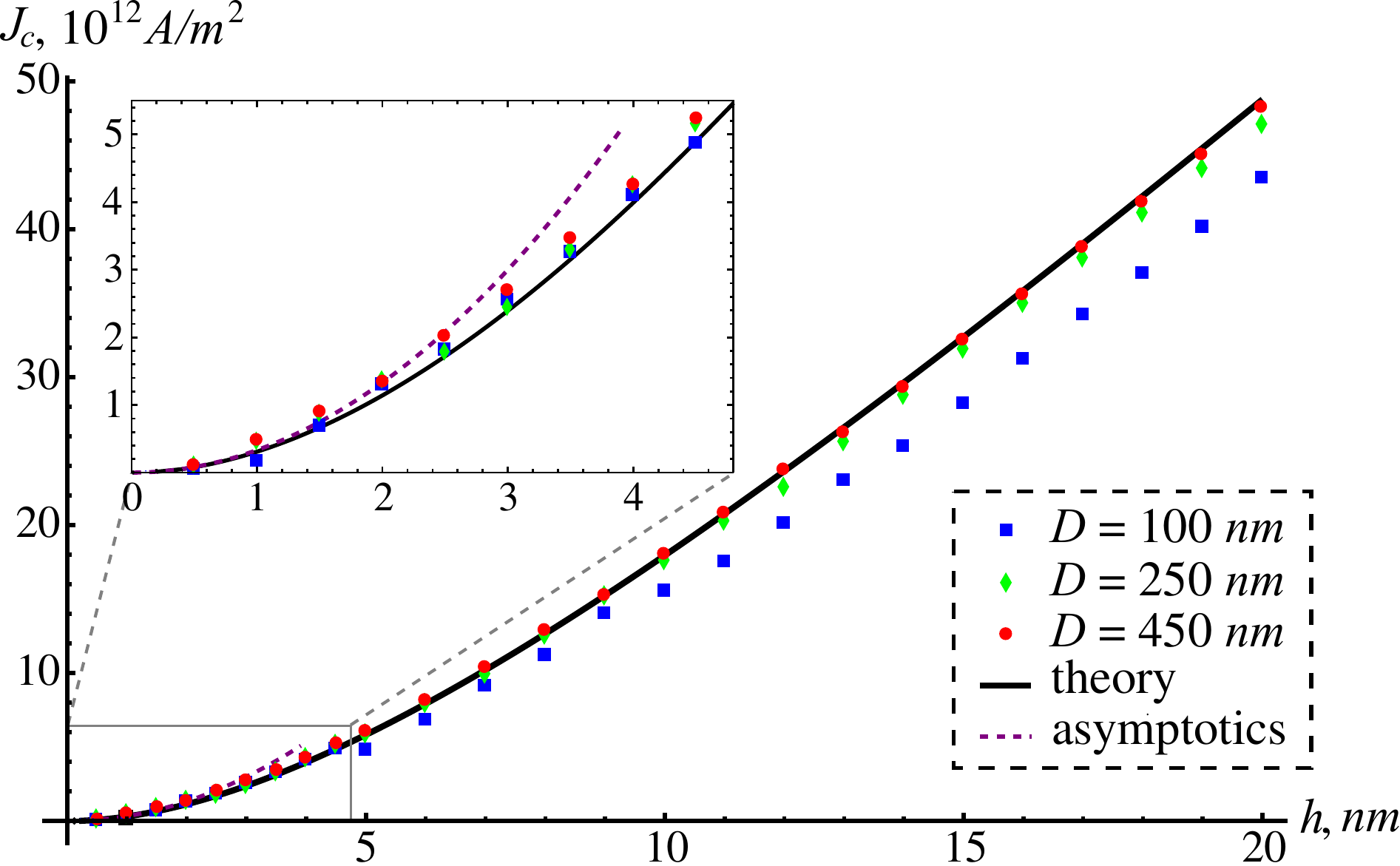}
\caption{Dependence of the saturation current $J_c$ on the film thickness. Solid line corresponds to the analytical solution obtained from \eqref{eq:Jc-def} and results of micromagnetic simulations (see Section~\ref{sec:simuls}) for different disk diameters $D$ are shown by markers. The dashed line demonstrates the parabolic asymptotic for $h\ll\ell$, see text.}\label{fig:Jc}
\end{figure}

For each thickness the curve $\varkappa=\tilde\varkappa(k)$, as well as the corresponding rescaled curve $J(k)$, separates stable and unstable regimes and has a maximum which determines the minimal current $J_c$, at which the saturated state remains stable. So the critical current at which the transition to saturation occurs can be determined as
\begin{equation} \label{eq:Jc-def}
J_c=\frac{2M_s^2e}{\eta\hbar}h\varkappa_c.
\end{equation}
As one can see from Fig.~\ref{fig:stab-diagramm} the saturation current $J_c$ increases with the increase of thickness. In detail this dependence is presented in the Fig.~\ref{fig:Jc}.

Using \eqref{eq:crit-current} and \eqref{eq:Jc-def} one can obtain the following asymptotic $J_c\approx h^2|e|M_s^2/(2\eta\hbar\ell)$ for $h\ll\ell$ and $J_c\approx h|e|M_s^2/(\eta\hbar)$ for $h\gg\ell$ though the last one is not achieved in the Fig.~\ref{fig:Jc} and it is beyond the limits of applicability of the Slonczewski torque in \eqref{eq:LLS}. The critical currents obtained using the micromagnetic simulations appears to be in a very good agreement with the theoretical curve. Since the present theory is constructed for a film of infinite lateral size the agreement between simulations and the theory is expectedly the best for samples whose thickness is much smaller than the planar size: $h\ll D$.



\section{Weakly nonlinear analysis}
\label{sec:nonlin}

In this section we prove the stability of structures with symmetry $C_4$ which appear in pre-saturation regime. We also show that these stable structures are square vortex-antivortex superlattices.  Thus, our analysis is based on the equation \eqref{eq:main-Four} where the nonlinear terms in the Hamiltonian \eqref{eq:all-energies} and in the driving function \eqref{eq:F-function} are taken into account.

The simplest way to describe the necessary symmetry is to restrict ourselves only with four wave-vectors $\vec k\in\left\{\vec K_\uparrow,\,\vec K_\rightarrow,\,\vec K_\downarrow,\,\vec K_\leftarrow\right\}$ in the wave-vectors space. Here we use the following notations
\begin{equation} \label{eq:4K}
\begin{split}
&\vec K_\uparrow=K(0,\,1),\qquad \vec K_\downarrow=K(0,\,-1),\\
&\vec K_\rightarrow=K(1,\,0),\qquad \vec K_\leftarrow=K(-1,\,0),
\end{split}
\end{equation}
where the amplitude $K$ is determined for given thickness from the linear analysis as following: $\varkappa_c=\tilde\varkappa(K)$, i.e. $K$ is the wave vector length which maximizes the dependence $\tilde\varkappa=\tilde\varkappa(k)$. It should be noted that since the Hamiltonian $\mathcal{E}$ contains $\hat\psi_{\vec k}$ as well as $\hat\psi_{-\vec k}$ then our model must contain pairs of vectors ($\vec K,\,-\vec K$). It means that only structures with even symmetry $C_{2n}$ are possible. We focus here on structures with symmetry $C_4$ in order to explain the results of the recent numerical experiments\cite{Volkov11}.

For the future analysis it is convenient to proceed to the following notations
\begin{equation} \label{eq:N-Phi-notation}
\hat\psi_{\vec K_\alpha}\equiv\sqrt{N_\alpha}e^{i\Phi_\alpha},
\end{equation}
where $\alpha\in\left\{\uparrow,\rightarrow,\downarrow,\leftarrow\right\}$. The value $N_\alpha$ in \eqref{eq:N-Phi-notation} has the meaning of the number of magnons with the corresponding wave vector. Substituting \eqref{eq:N-Phi-notation} into the equation of motion \eqref{eq:main-Four} we obtain the set of eight equations
\begin{subequations}\label{eq:eq-motion-N-Phi}
\begin{align} \label{eq:eq-motion-N}
\dot N_\uparrow &= -\frac{\partial\mathcal{E}}{\partial\Phi_\uparrow}-2\varkappa N_\uparrow\left[1-\frac{\sum_\alpha N_\alpha-\frac12N_\uparrow}{\Lambda^2\mathcal{N}_{xy}}\right]\\ \nonumber
&+\frac{2\varkappa}{\Lambda^2\mathcal{N}_{xy}}\sqrt{N_\uparrow N_\downarrow N_\rightarrow N_\leftarrow}\cos(\Phi_\updownarrow-\Phi_\leftrightarrow),\\
\label{eq:eq-motion-Phi}
\dot\Phi_\uparrow &= \frac{\partial\mathcal{E}}{\partial N_\uparrow}-\frac{\varkappa}{\Lambda^2\mathcal{N}_{xy}}\frac{\sqrt{N_\downarrow N_\rightarrow N_\leftarrow}}{\sqrt{N_\uparrow}}\sin(\Phi_\updownarrow-\Phi_\leftrightarrow),
\end{align}
\end{subequations}
where the other three pairs of equations can be obtained by three-time successive rotations of all subscripts by the angle $\pi/2$, and we introduced the notations $\Phi_\updownarrow=\Phi_\uparrow+\Phi_\downarrow$ and $\Phi_\leftrightarrow=\Phi_\rightarrow+\Phi_\leftarrow$ for the sake of simplicity. The Hamiltonian in ``$N-\Phi$''-notation being presented as a sum of linear and nonlinear parts is the following
\begin{subequations} \label{eq:Hamiltonian-N-Phi}
\begin{equation} \label{eq:Hamiltonian-N-Phi-1}
\mathcal{E}=\mathcal{E}^0+\mathcal{E}^\mathrm{nl},
\end{equation}
where the linear part reads
\begin{equation} \label{eq:Hamiltonian-N-Phi-lin}
\begin{split}
\mathcal{E}^0 &= \left[\ell^2K^2-1+\frac{g_1}{2}\right]\left(N_\updownarrow+N_\leftrightarrow\right)\\
 &-g_1\left[\sqrt{N_\Updownarrow}\cos(\Phi_\updownarrow)-\sqrt{N_\Leftrightarrow}\cos(\Phi_\leftrightarrow)\right],
\end{split}
\end{equation}
and the fourth-order nonlinearity has the following form
\begin{equation} \label{eq:Hamiltonian-N-Phi-nl}
\begin{split}
&\mathcal{E}^\mathrm{nl}=\frac{2}{\mathcal{N}_{xy}}\Biggl\{-\frac{1}{4}\left[\ell^2K^2-1+\frac{g_1}{2}\right]\sum\limits_\alpha N_\alpha^2\\ &+\left[\ell^2K^2+1-\frac{g_1+g_2}{2}\right] (N_\Updownarrow+N_\Leftrightarrow)\\
&+2\left[\ell^2K^2+1-g_{\sqrt2}-\frac{g_1}{2}\right]\sqrt{N_\Updownarrow N_\Leftrightarrow}\\
&\times \cos(\Phi_\updownarrow-\Phi_\leftrightarrow)+\left[1-\frac{g_1+g_{\sqrt2}}{2}\right]N_\updownarrow N_\leftrightarrow\\
&+\frac{g_1}{4}\biggl[\sqrt{N_\Updownarrow}\left(\frac32N_\updownarrow+N_\leftrightarrow\right)\cos \Phi_\updownarrow -\\
&-\sqrt{N_\Leftrightarrow}\left(\frac32N_\leftrightarrow+N_\updownarrow\right)\cos \Phi_\leftrightarrow\biggr]\Biggr\}.
\end{split}
\end{equation}
\end{subequations}
Here we used the analogous notations $N_\updownarrow=N_\uparrow+N_\downarrow$, $N_\leftrightarrow=N_\rightarrow+N_\leftarrow$, $N_\Updownarrow=N_\uparrow N_\downarrow$, $N_\Leftrightarrow=N_\rightarrow N_\leftarrow$ and $g_\xi\equiv g(\xi Kh)$ to shorten the expressions.

Using \eqref{eq:eq-motion-N} and \eqref{eq:Hamiltonian-N-Phi} one can show that
\begin{equation} \label{eq:dt-N-Phi}
\frac{\mathrm{d}}{\mathrm{d}t}(N_\uparrow-N_\downarrow)=-2\varkappa(N_\uparrow-N_\downarrow)\left[1-\frac{N_\updownarrow +2N_\leftrightarrow}{2\Lambda^2\mathcal{N}_{xy}}\right]
\end{equation}
with the corresponding equation for the subscripts rotated by $\pi/2$. Taking into account that $\varkappa>0$ we conclude from the Eq.~\eqref{eq:dt-N-Phi} that after period of time $\Delta\tau=1/(2\varkappa)$ the system achieves a stationary regime with $N_\uparrow=N_\downarrow$ and $N_\rightarrow=N_\leftarrow$. Consideration of these conditions in the stationary form of system \eqref{eq:eq-motion-N-Phi} leads to possibility of a solution which satisfy the following conditions
\begin{equation} \label{eq:N-Phi-cond}
\begin{split}
&N_\uparrow=N_\downarrow=N_\rightarrow=N_\leftarrow=N,\\
&\Phi_\uparrow+\Phi_\downarrow=\pi+\Phi_\rightarrow+\Phi_\leftarrow=\Phi.
\end{split}
\end{equation}
Under the condition \eqref{eq:N-Phi-cond} all four pairs of stationary equations of motion \eqref{eq:eq-motion-N-Phi} become identical and they obtain the following form
\begin{subequations}\label{eq:N-Phi-final}
\begin{align}
\label{eq:N}&\sin\Phi \left(1-\frac52\mathscr{N}\right)g_1=-2\varkappa\left(1-\frac52\frac{\mathscr{N}}{\Lambda^2}\right),\\
\label{eq:cosPhi}&\cos\Phi\left(1-5\mathscr{N}\right)\frac{g_1}{2}=\ell^2K^2-1+\frac{g_1}{2}\\ \nonumber
&-\mathscr{N}\left(\ell^2K^2-5+\frac52g_1+g_2\right),
\end{align}
\end{subequations}
where $\mathscr{N}=N/\mathcal{N}_{xy}$ is density of the magnons, and the energy density obtained from the Hamiltonian \eqref{eq:Hamiltonian-N-Phi} reads
\begin{equation} \label{eq:Hamiltonian-simple}
\begin{split}
&\frac{\mathcal{E}}{\mathcal{N}_{xy}}=2\mathscr{N}\left[2\left(\ell^2K^2-1\right)+g_1(1-\cos\Phi)\right]\\
&+2\mathscr{N}^2\left[-\ell^2K^2+5-g_2-\frac52g_1(1-\cos\Phi)\right].
\end{split}
\end{equation}

\begin{figure}
\includegraphics[width=\columnwidth]{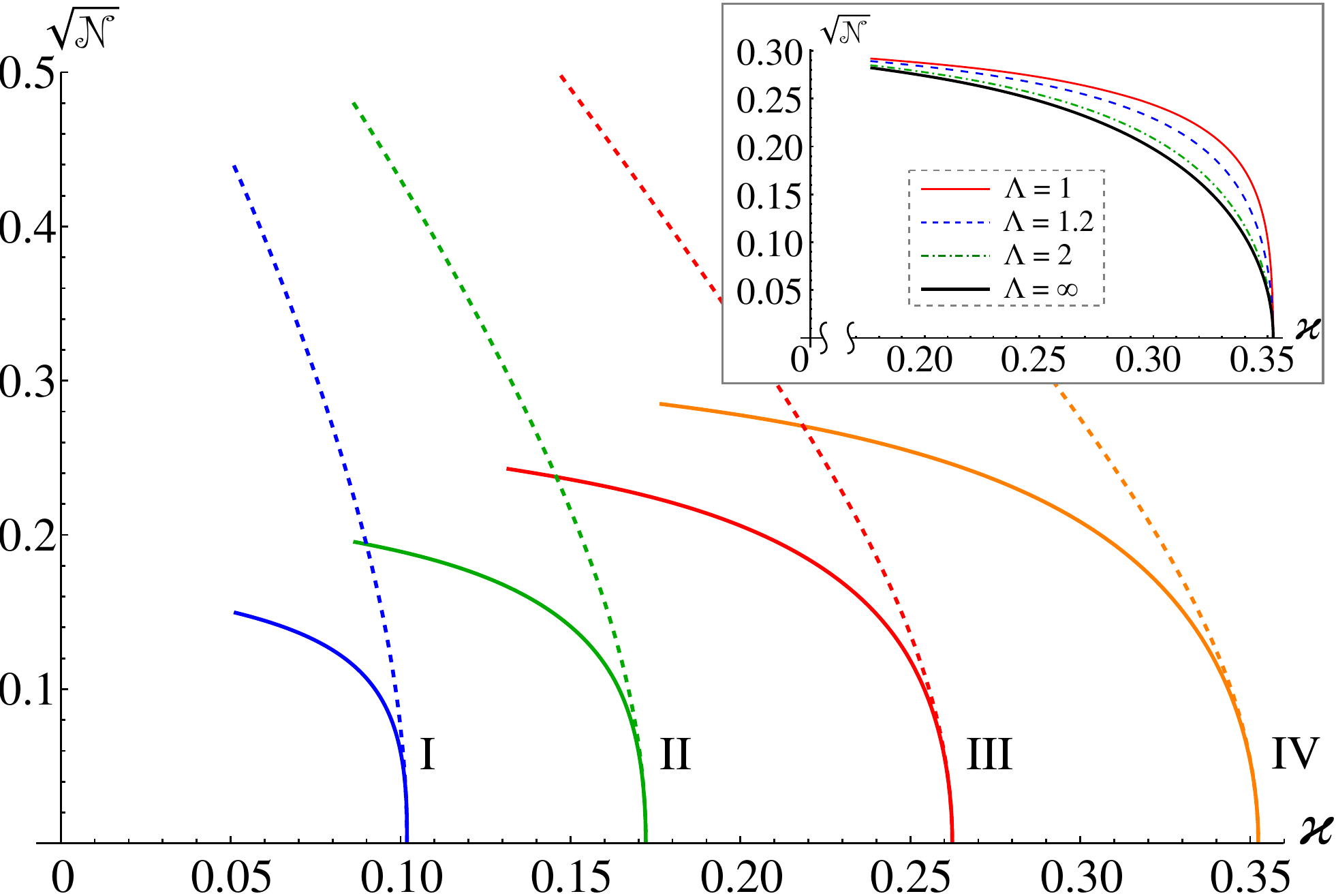}
\caption{Magnons density as function of the normalized current for different thicknesses (in units of $\ell$): I -- 0.5, II -- 1, III -- 2, IV -- 4 and $\Lambda=2$ for all thicknesses. Solid lines show the exact numerical solutions of the system \eqref{eq:N-Phi-final} and dashed lines correspond to the approximation \eqref{eq:N-cosPhi-final}. The inset demonstrates the weakness of influence of the parameter $\Lambda$ on the exact solution, the data corresponds to the thickness $h=4\ell$. Each of the plots is built for the range $[\varkappa_c/2,\,\varkappa_c]$.}\label{fig:N_vs_kappa}
\end{figure}

Excluding $\Phi$ from \eqref{eq:N-Phi-final} and taking into account that $\mathscr{N}\ll1$ one obtains
\begin{subequations}\label{eq:N-cosPhi-final}
\begin{align}
\label{eq:N-final}&\mathscr{N}\approx\frac{\varkappa_c^2(h)-\varkappa^2}{\mathfrak{GF}+5\varkappa^2\left(1-\frac{1}{\Lambda^2}\right)},
\end{align}
and then using \eqref{eq:cosPhi} one can estimate
\begin{align}
\label{eq:cosPhi-final}&\cos\Phi\approx\frac{\mathfrak{F}}{g(Kh)}\sqrt{1+4\frac{\mathfrak{G}}{\mathfrak{F}}\mathscr{N}}.
\end{align}
\end{subequations}
Here we introduced the following thickness dependent functions $\mathfrak{F}=2[K^2\ell^2-1+\frac12g(Kh)]$ and $\mathfrak{G}=[4K^2\ell^2-g(2Kh)]$. Magnon density \eqref{eq:N-final} as well as the corresponding exact solutions of Eqs. \eqref{eq:N-Phi-final} are shown in the Fig.~\ref{fig:N_vs_kappa}. As one can see, the approximation \eqref{eq:N-final} is satisfactory near the instability threshold.

\begin{figure}
\includegraphics[width=0.85\columnwidth]{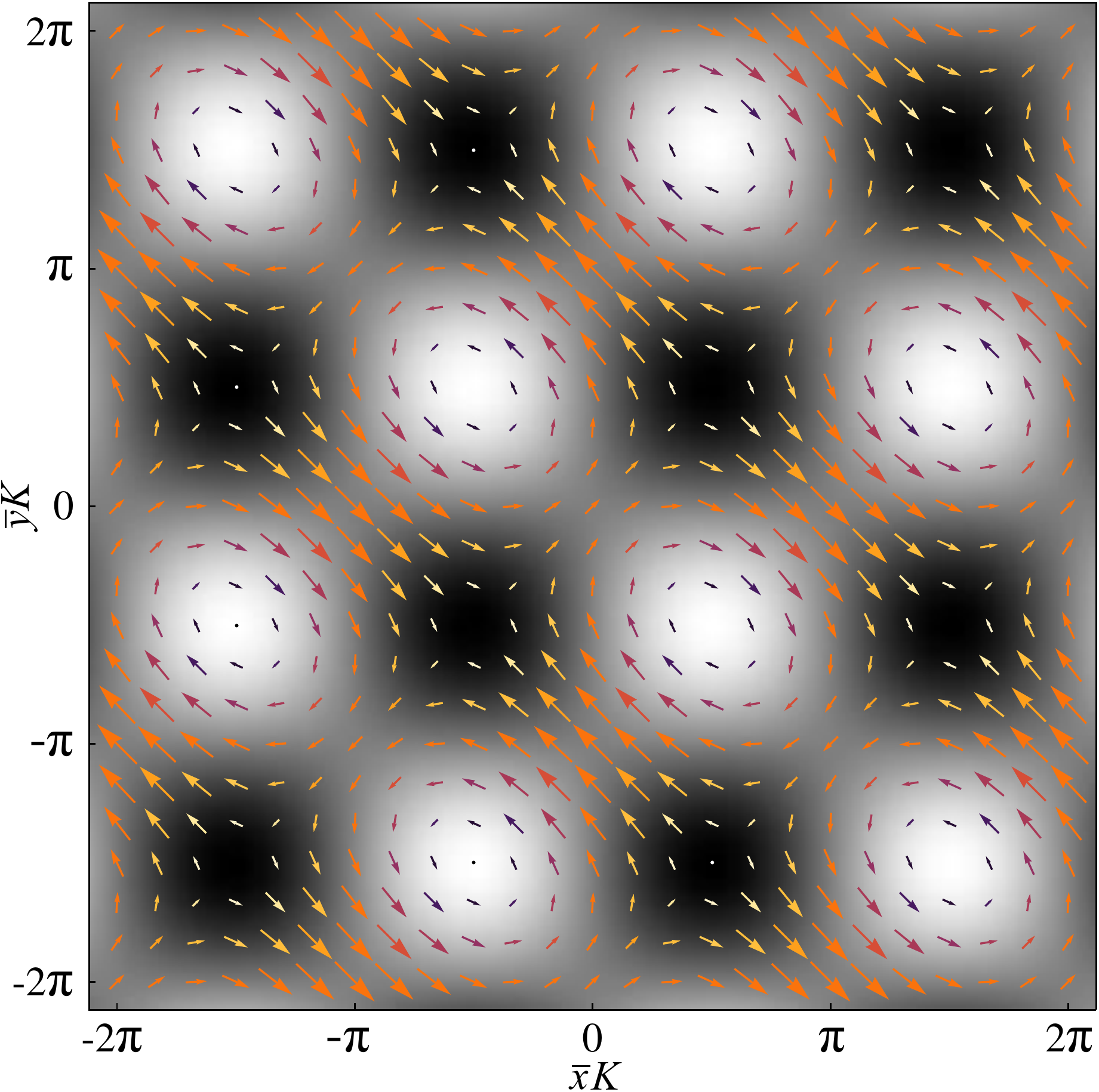}
\caption{The analytically obtained vortex-antivortex superlattice. Arrows show distribution of magnetization \eqref{eq:mx-my-4-wave} and the corresponding topological density \eqref{eq:top-den} is shown by gray tones. The figure is built for the case $h=4\ell$ and $\varkappa=0.65\varkappa_c$ and $\Lambda=2$ (the required value of $\Phi$ was determined from \eqref{eq:N-Phi-final} for the mentioned parameters).}\label{fig:crystal_theor}
\end{figure}

Thus we have proved the possibility of a stationary structure which is described by the four-wave Ansatz \eqref{eq:4K} and \eqref{eq:N-Phi-notation} in the pre-saturated regime. At the same time one can see that the parameter $\Lambda$ does not influence considerably the system behavior, see the inset in the Fig.~\ref{fig:N_vs_kappa}.

The linear stability analysis for the system \eqref{eq:eq-motion-N-Phi}shows that the stationary solution \eqref{eq:N-cosPhi-final} is stable in the close vicinity of the critical current
$$
0<\frac{\varkappa_c-\varkappa}{\varkappa_c}\ll1,
$$
 see Appendix~\ref{app:stability} for details.

Let us now see how the mentioned structure looks like. From \eqref{eq:four-inv} one can obtain the following expression for $\psi$-function
\begin{equation} \label{eq:psi-4wave}
\psi_{\vec n}=\frac{1}{\sqrt{\mathcal{N}_{xy}}}\sum\limits_\alpha\sqrt{N_\alpha}e^{i(\Phi_\alpha+\vec K_\alpha\cdot\vec n)}.
\end{equation}
Varying parameters $N_\alpha$ and $\Phi_\alpha$ one can obtain a wide range of different structures from \eqref{eq:psi-4wave} but under the conditions \eqref{eq:N-Phi-cond} the expression \eqref{eq:psi-4wave} results exactly the square vortex-antivortex superlattice. Indeed, substituting \eqref{eq:psi-4wave} into \eqref{eq:mxmymz} with taking into account the conditions \eqref{eq:N-Phi-cond} we obtain in the linear approximation
\begin{equation} \label{eq:mx-my-4-wave}
\begin{split}
&m_x\approx2\sqrt{2\mathscr{N}}\left[\cos(K\bar x)\sin\frac{\Phi}{2}+\cos(K\bar y)\cos\frac{\Phi}{2}\right],\\
&m_y\approx2\sqrt{2\mathscr{N}}\left[\cos(K\bar x)\cos\frac{\Phi}{2}+\cos(K\bar y)\sin\frac{\Phi}{2}\right],\\
&m_z\approx1,
\end{split}
\end{equation}
were the following shift of the coordinate origin was performed: $\bar x=x+(\Phi_\rightarrow-\Phi_\leftarrow)/(2K)$ and $\bar y=y+(\Phi_\uparrow-\Phi_\downarrow)/(2K)$. Magnetization distribution which corresponds to \eqref{eq:mx-my-4-wave} for certain parameters is shown in the Fig.~\ref{fig:crystal_theor} by arrows. The topological properties of the system can be characterized by the topological density\cite{Papanicolaou91} (or scalar chirality density\cite{Solenov12})  $\upsilon=[\partial_x\vec m\times\partial_y\vec m]\cdot\vec m$. The topological density which corresponds to \eqref{eq:mx-my-4-wave} reads
\begin{equation} \label{eq:top-den}
\upsilon=-8K^2\mathscr{N}\cos\Phi\sin(K\bar x)\sin(K\bar y).
\end{equation}
The distribution of \eqref{eq:top-den} is shown in the Fig.~\ref{fig:crystal_theor} by gray tones. It resembles chirality waves in Kondo magnets.\cite{Solenov12}

\begin{figure}
\includegraphics[width=0.9\columnwidth]{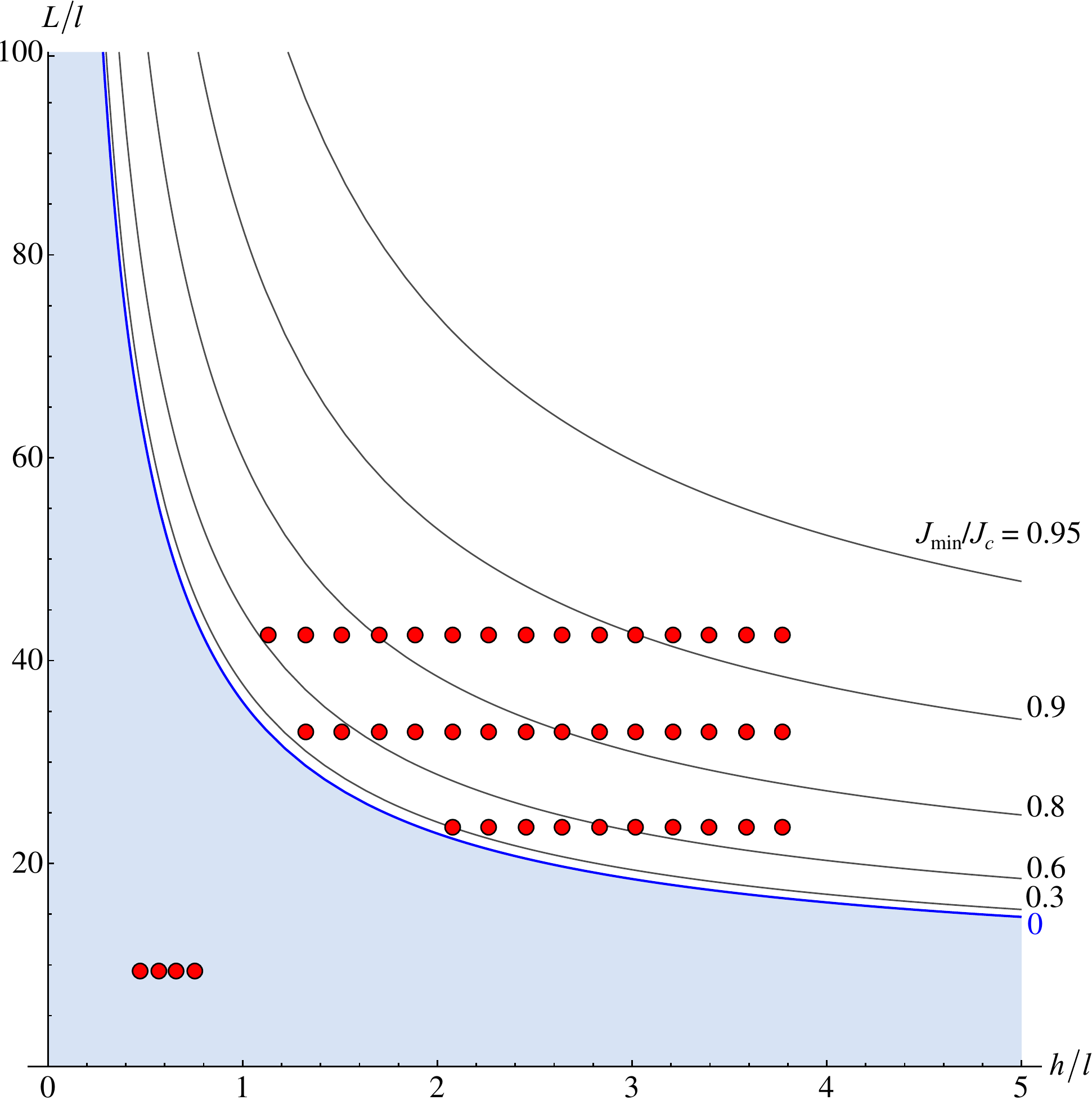}
\caption{The diagram which determines the range $J\in[J_\mathrm{min},\,J_c]$ for the given geometry sizes $(h,\,L)$, where the model \eqref{eq:4K}-\eqref{eq:N-Phi-notation} is applicable. In shaded region the model works for any currents $J<J_c$. Points correspond to the disks, where the superlattices were observed via micromagnetic simulations.
}\label{fig:applicability}
\end{figure}

Though the model \eqref{eq:4K}-\eqref{eq:N-Phi-notation} results the superlattice very similar to one which is observed in the numerical experiment, one should point out domain of applicability of this model. In \eqref{eq:4K}--\eqref{eq:N-Phi-notation} we use only the critical value of the wave-vector $K$ instead the whole possible wave-vectors in the range $\left[K',\,K''\right]$, where $K'$ and $K''$ bound the instability domain for the given current value, see Fig.~\ref{fig:stab-diagramm}. We can restrict ourselves with the single $K$ if only size of the system is small enough:
\begin{equation} \label{eq:applicability}
2\pi/L>K''-K'.
\end{equation}
Since the domain $\Delta K=K''-K'$ increases with the current decreasing (see Fig.~\ref{fig:stab-diagramm}) the condition \eqref{eq:applicability} is equivalent to $J_\mathrm{min}<J<J_c$, where $J_\mathrm{min}$ is the minimal current at which the condition \eqref{eq:applicability}. For each of the values of thickness $h$ and radius $L$ one can calculate the value of $J_\mathrm{min}$ using \eqref{eq:crit-current} and \eqref{eq:applicability}, the resulting diagram is shown in the Fig.~\ref{fig:applicability}. Since for the small thicknesses $\Delta K\ll1$ the model \eqref{eq:4K}--\eqref{eq:N-Phi-notation} can be used for wide range of currents.

\section{Micromagnetic simulations}
\label{sec:simuls}

To investigate numerically the process of magnetic film saturation under the influence of spin-polarized current we used full scale OOMMF \cite{OOMMF} micromagnetic simulations. All simulations were performed for disk shaped nanoparticles with material parameters of permalloy: saturation magnetization $M_S=8.6 \times 10^5$ A/m, exchange constant $A=13 \times 10^{-12}$ J/m, and the anisotropy was neglected. The damping was neglected, because, as it was shown in Section~\ref{sec:Harmonic}, the spin-polarized current plays role of an effective damping. The mesh cell was chosen to be $3 \times 3 \times h$ nm. The current parameters $\eta = 0.4$, and $\Lambda = 2$ were the same for all simulations, except some cases mentioned in the text bellow.

In the first stage we obtained the dependence of saturation current $J_c$ on the sample thickness. For this numerical experiment we chose the nanodisks with three different diameters $D=100$, 250 and 450 nm respectivelly, and thickness of each of the particles was varied from 0.5 nm  to 20 nm. As the initial state for a simulation the ground state of the particle was chosen: uniform magnetization within the sample plane for thin disks ($h<5$ nm) and vortex state for thicker ones. The spin-current was increased until the saturation was achieved. As a criterion of the saturation we used the relation $M_z/M_s>0.9999$, where $M_z$ is the total magnetization along the current direction. The resulting dependence $J_c(h)$ is shown in the Fig.~\ref{fig:Jc} by markers. As one can see, for disks with the small aspect ratio the micromagnetic simulations confirm the analytical results with a high accuracy. The slight deviation from the theoretically predicted curve is observed for the case of small disks (see $D=100$ nm in the Fig.~\ref{fig:Jc}). This is because the presented theory is build for the case of an infinite film what corresponds to zero aspect ratio.

\begin{figure}
\includegraphics[width=\columnwidth]{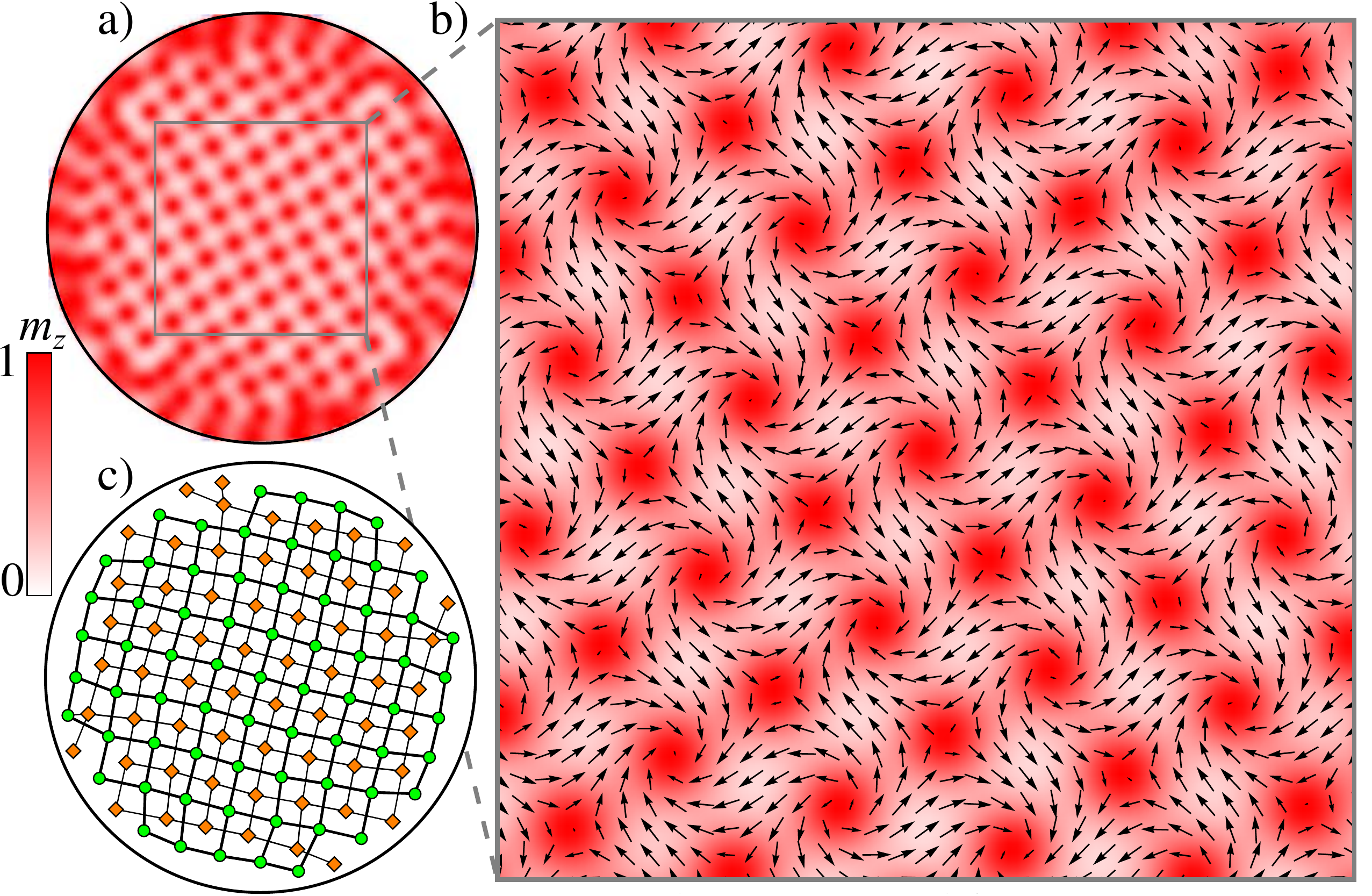}
\caption{The superlattice structure obtained using simulations in disk with diameter $D=350$ nm and thickness $h=20$ nm under influence of the current $J=32\times10^{12}\,A/m^2$. Inset a) shows the out-of-plane structure of the superlattice, inset b) demonstrates in details the magnetization of central part of the the disk: arrows correspond to the in-plane magnetization distribution and out-of-plane component $m_z$ is shown by color. The superlattice properly is shown in the inset c): positions of vortices and antivortices are shown by disks and rhombuses respectively.}\label{fig:crystal}
\end{figure}

\begin{figure}
\includegraphics[width=\columnwidth]{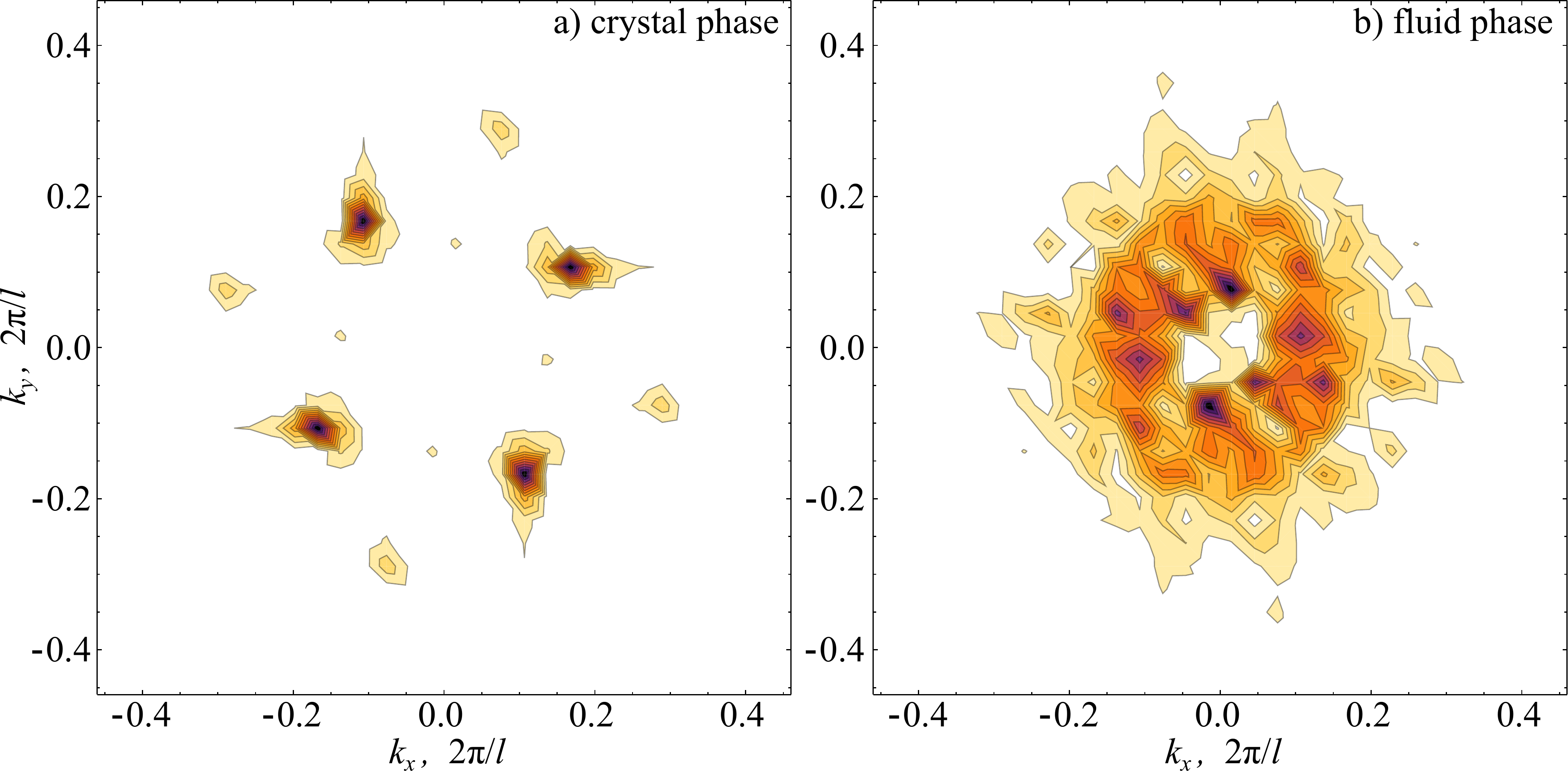}
\caption{Two-dimensional Fourier spectrums of the crystal a), and fluid b) structures. Inset a) shows the Fourier transform of the function $m_z(x,y)-\langle m_z\rangle$ for the case of the crystal structure shown in the Fig.~\ref{fig:crystal}~b), where $\langle m_z\rangle$ is the averaged $m_z$-component. And the inset b) corresponds to a fluid structure obtained for current $J=25\times10^{12}\,A/m^2$, the other parameters are the same as in the Fig.~\ref{fig:crireruim}.}\label{fig:fourier}
\end{figure}

To study the magnetization dynamics in regime $J\lesssim J_c$ we used a disk with diameter $D=350$ nm and thickness $h=20$ nm. A spin-current of the certain density was sharply applied to this nanodisk, which initially was in the vortex ground state. After a few nanoseconds a slowly rotating superlattice was formed for case of the current value close to the saturation, see the Fig.~\ref{fig:crystal}, or a fluid-like dynamics of locally ordered vortex-antivortex media was observed for cases of lower currents, see Fig.~2 in the Ref.~\onlinecite{Volkov11}. The Fourier spectrums of the typical crystal and fluid structures are compared in the Fig.~\ref{fig:fourier}. Accordingly to the Fig.~\ref{fig:fourier}a) the superlattice is square one.

To separate crystal and fluid phases and to study their properties we performed a series of simulations for a range of currents $J\in[J_c/2,\;J_c]$ with the current step $\Delta J=0.5\times10^{12}\,A/m^2$. For a certain value of the current the magnetization dynamics was simulated for 30 ns. Starting from the time moment 2 ns we saved the magnetization distribution with the time step 0.2 ns.  For each of the saved in this way magnetization snap shots we found coordinates of all particles (vortices and antivortices) using the method\cite{Hertel06} of  intersection of isolines  $m_x=0$ and $m_y=0$. To distinguish vortices from antivortices the winding number of each of the particles was calculated as circulation on small circumference centered on the particle position. Then for each of the vortices the distances to the nearest four antivortices were found (on this stage to avoid the boundary influence we consider only vortices distanced from the disk center less then a half of the disk radius). Then the histogram of the distribution of all obtained vortex-antivortex distances was build for a certain magnetization snap shot, and finally we build the averaged histogram based on all magnetization snap shots for a certain current value. Two examples of these averaged histograms are shown in the left column of the Fig.~\ref{fig:crireruim}.  The obtained histograms can be well fitted by the Gaussian $f(x)\propto \exp\left[-(x-x_0)^2/\sigma^2\right]$, where $f(x)$ is number of the vortex-antivortex distances which are in the interval $[x,\;x+\Delta x]$ with $\Delta x=1$ nm being the width of the histogram bin.

\begin{figure}
\includegraphics[width=0.85\columnwidth]{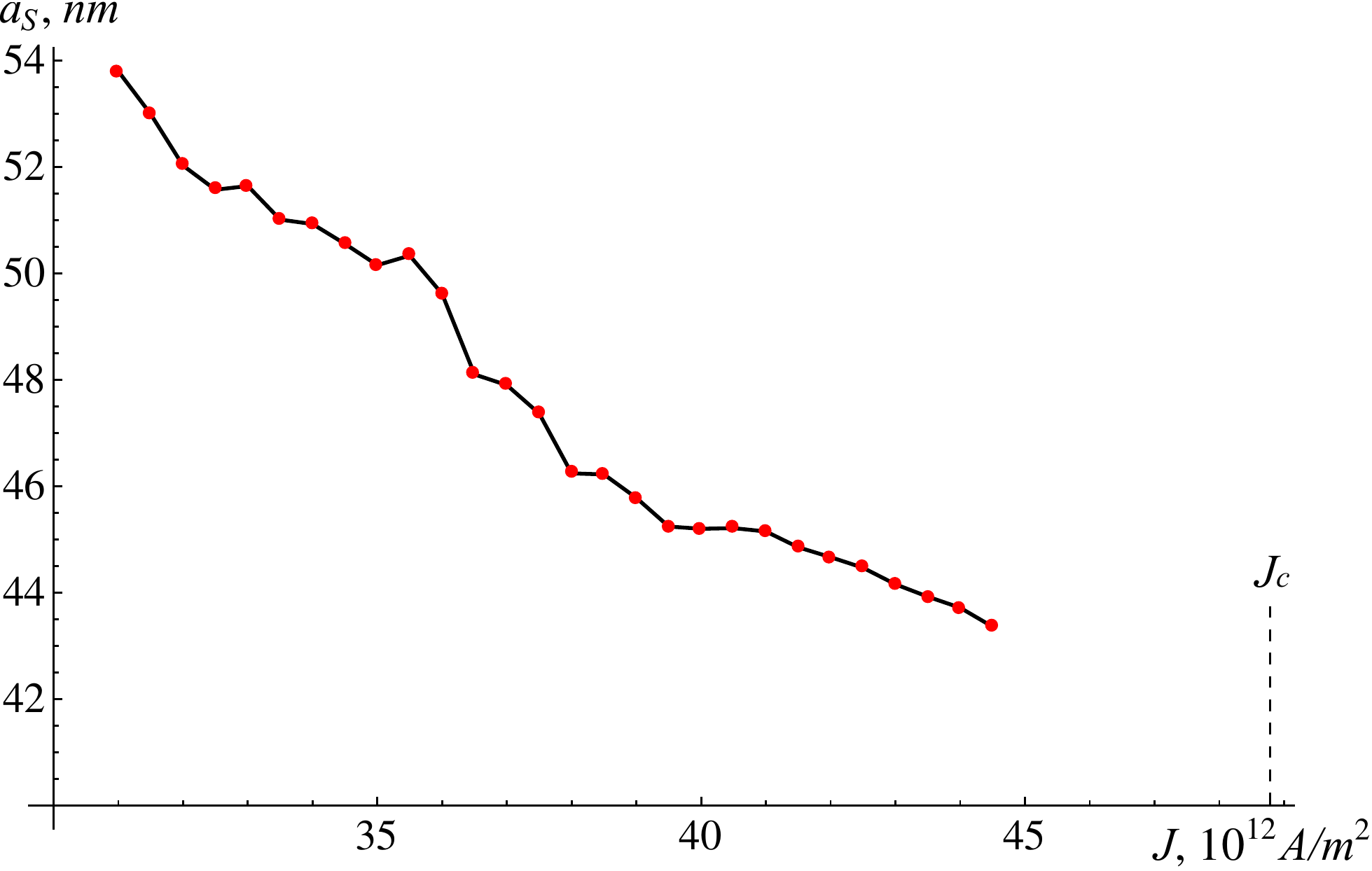}
\caption{Dependence of the superlattice constant $a_S$ on the applied current for permalloy disk with diameter $D=350\,nm$ and thickness 20nm.  Data were obtained using micromagnetic simulations.}\label{fig:a-vs-j}
\end{figure}

For the crystal phase the superlattice constant was considered to be $a_s=2 x_0$. We found that the superlattice constant slightly decreases with the current increasing, this dependence is shown in the Fig.~\ref{fig:a-vs-j}. However we were not able to determine $a_s$ very close to the saturation current because the components of magnetization $m_x$ and $m_y$ become vanishingly small.  The obtained dependence $a_s(J)$ appears to be not very smooth because of the stress in the superlattice due to presence of the boundary.

One of the important characteristics, which can be extracted from histograms is the $\sigma(J)$--dependence. It gives a possibility quantitatively separates fluid phase and crystal one: in crystals the value of $\sigma$ is small (about a few nanometers) and it is weakly dependent on the current $J$; in the fluid phase the value of $\sigma$ increases fast with the current decreasing. To determine the critical current $J_{fc}$ of transition between fluid and crystal phases we fit the numerically obtained dependence $\sigma(J)$ by the function $\sigma=(a J+b)\theta(-J+J_{fc})+(a J_{fc}+b)\theta(J-J_{fc})$ with $\theta(x)$ being the Heaviside step function and $a,\,b$ being the fitting parameters, see Fig.~\ref{fig:crireruim}.

Accordingly to the linear analysis (Section~\ref{sec:Harmonic}) the parameter $\Lambda$ does not influence the saturation current $J_c$, and accordingly to the weakly nonlinear analysis of the pre-saturated regime (Section~\ref{sec:nonlin}) the parameter $\Lambda$ influences very weakly on the dynamics of the vortex-antivortex superlattice. Our theory is not able to describe the transition between fluid and crystal phases, but using the simulations and the methods described above we found that the current $J_{fc}$, and consequently the current range $[J_{fc},\,J_{c}]$ of the crystal phase existence, depend on the parameter $\Lambda$, see the Fig.~\ref{fig:J_f_c4_from_Lambda}. We found out that dependence $J_{fc}(\Lambda)$ can be well fitted by the function
\begin{equation} \label{eq:Jfc-vs-Lambda}
J_{fc}=\frac{\beta}{\sqrt{\Lambda^2-1}}+J_{fc}^0,
\end{equation}
with $\beta\approx7.78\times10^{12}\,A/m^2$ and $J_{fc}^0\approx25.92\times10^{12}\,A/m^2$ being the critical current of the phase transition for case $\Lambda\rightarrow\infty$. For the case $\Lambda=1$ the superlattice is not formed, only the fluid-like dynamics is observed. Also the current region of the crystal phase quickly shortens when the parameter $\Lambda$ is reduced to 1, but for $\Lambda>4$ the the crystal region is approximately constant.

It should be noted that we do not consider here the ``gas" phase and the rarefied patterns which were observed in Ref.~\onlinecite{Volkov11} at lower currents $J\ll J_{c}$.

\begin{figure}
\includegraphics[width=\columnwidth]{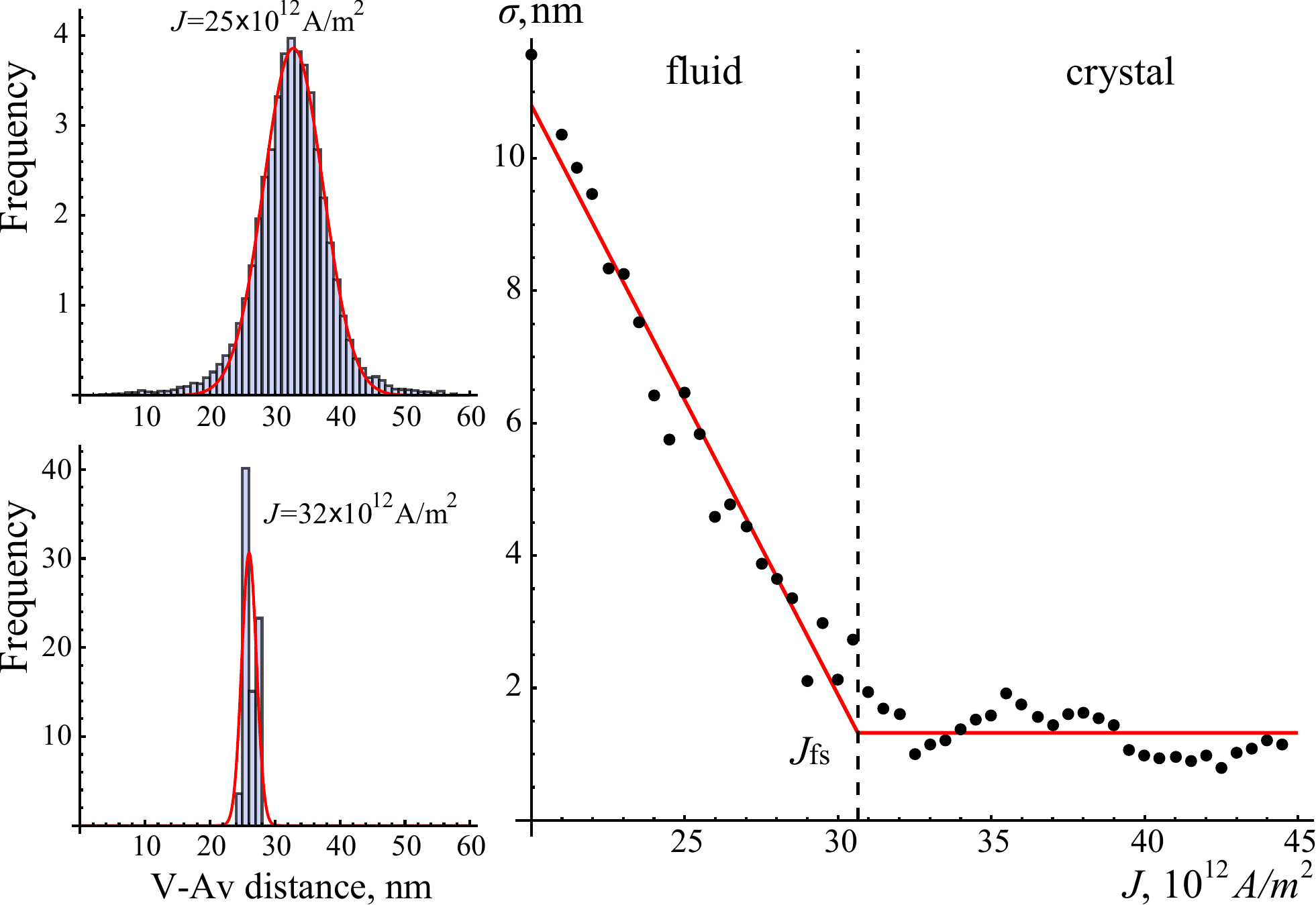}
\caption{The criterium of separation of fluid and crystal phases. In the left column the distributions of the distances between the nearest vortices and antivortices are presented, solid line shows the Gaussian approximation. The upper and lower histograms correspond to the typical fluid-like  and crystal-like structures respectively. The right plot demonstrates dependence of half-width of the mentioned distributions on the applied current. All data are obtained from simulations for disk with $D=350$ nm, $h=20$ nm and $\Lambda=2$.}\label{fig:crireruim}
\end{figure}


\begin{figure}
\includegraphics[width=\columnwidth]{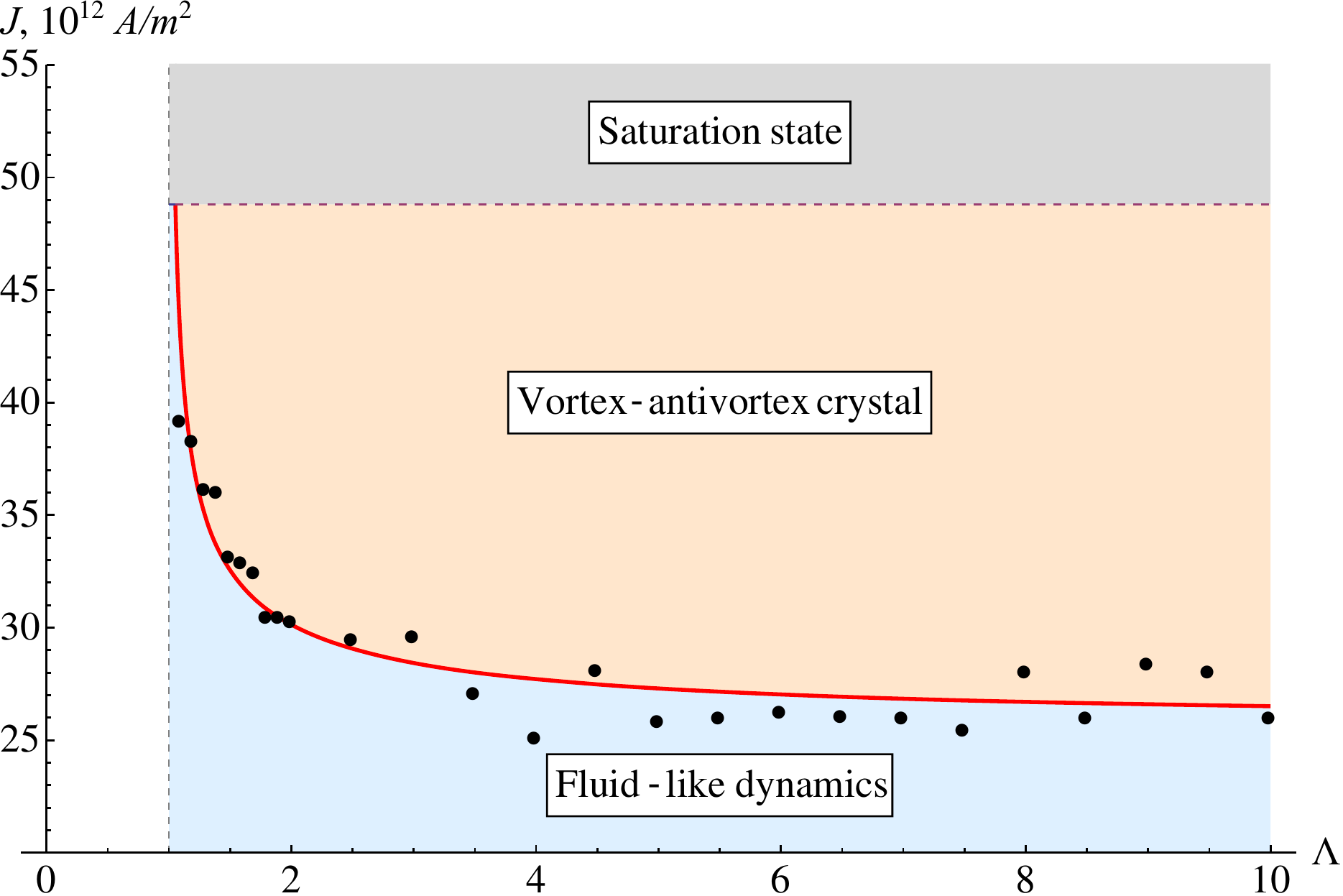}
\caption{Phase diagram of the pre-saturated magnetic film with thickness $h=20$ nm. The transition current $J_{fc}$, obtained from the simulation data (see Section~\ref{sec:simuls} and Fig.~\ref{fig:crireruim}) is shown by points and the corresponding fitting \eqref{eq:Jfc-vs-Lambda} is shown by the solid line.}\label{fig:J_f_c4_from_Lambda}
\end{figure}


\section{Conclusions}
\label{sec:conclusion}
We studied theoretically the process of vortex-antivortex pattern formation  in thin ferromagnetic films  under the action of strong transversally spin-polarized current. We show that there exists a critical (or saturation) current  $J_c$ above which the film goes to a saturated state with all magnetic moments directed perpendicularly to the film plane. The critical current strongly depends on the sample thickness and it is practically independent on the lateral size of the magnet. The saturation current increases with the thickness increasing following squared law for thin samples and linear one for thick samples. We demonstrate that the stable regular structures with symmetry $C_4$ can appear in pre-saturated regime and we show that these structures are square vortex-antivortex superlattices. Spatial period of the superlattice slightly decreases with the current increasing. The micromagnetic simulations confirm our analytical results with a high accuracy. Using the simulations we describe the melting of the vortex crystal with the current decrease.

We show that parameter $\Lambda$ which controls the spin-transfer torque efficiency, does not modify significantly neither the saturation current $J_c$ nor the dynamics of the vortex-antivortex superlattice. In contrast to this, the critical current $J_{fc}$  which gives the boundary between the fluid phase and the crystal one is very sensitive to the spin-torque  efficiency parameter $\Lambda$: the interval of the crystal phase existence $\left[J_{fc},\,J_{c}\right]$  contracts when $\Lambda\rightarrow1$ and it is constant for $\Lambda\gg1$.

\appendix

%

\section{Hamiltonian in the reciprocal space}
\label{app:Energies}

Here we calculate the magnetic energy in the wave--vector space limiting ourselves by the 4-th order nonlinearity.

\subsection{Exchange energy}
\label{app:Energies-Eex}

Let us consider first the exchange energy. Substituting the magnetization \eqref{eq:mxmymz} written in terms of $\psi$ into the general expression \eqref{eq:Eex} one can write the total exchange energy in form $E_\mathrm{ex}\approx E_\mathrm{ex}^0+E_\mathrm{ex}^\mathrm{nl}$, where the linear part reads
\begin{equation*}
\begin{split}
E_\mathrm{ex}^0 &=-\mathcal{S}^2\mathcal{N}_z\sum\limits_{\vec n,\vec l\ne\vec0}\mathcal{J}_{\vec l}\biggl[\psi_{\vec n}\psi_{\vec n+\vec l}^*\\
&-\frac12\left(|\psi_{\vec n}|^2+|\psi_{\vec n+\vec l}|^2\right)+\text{c.c.}\biggr]
\end{split}
\end{equation*}
and the corresponding nonlinear part takes the form
\begin{equation*}
\begin{split}
E_\mathrm{ex}^{\mathrm{{nl}}}&=-\frac{\mathcal{S}^2\mathcal{N}_z}{4}\sum\limits_{\vec n,\vec l\ne\vec0}\mathcal{J}_{\vec l}\biggl[\psi_{\vec n}\psi_{\vec n+\vec l}^*\left(|\psi_{\vec n}|^2+|\psi_{\vec n+\vec l}|^2\right)\\
&-2|\psi_{\vec n}|^2|\psi_{\vec n+\vec l}|^2+\text{c.c.}\biggl].
\end{split}
\end{equation*}
Let use perform the Fourier transform \eqref{eq:four-inv} with account of the orthogonality condition \eqref{eq:orth-cond}, which results in
\begin{align*}
&E_\mathrm{ex}^0=2\mathcal{S}^2\mathcal{N}_z\sum\limits_{\vec l}\mathcal{J}_{\vec l}\sum\limits_{\vec k}|\hat\psi_{\vec k}|^2\left(1-e^{i\vec k\cdot\vec l}\right),\\
&E_\mathrm{ex}^{\mathrm{{nl}}}=\frac{\mathcal{S}^2\mathcal{N}_z}{4\mathcal{N}_{xy}}\sum\limits_{\vec l}\mathcal{J}_{\vec l}\!\!\!\sum\limits_{\vec k_1\vec k_2\vec k_3\vec k_4}\!\Biggl[\hat\psi_{\vec k_1}\hat\psi_{\vec k_2}^*\hat\psi_{\vec k_3}\hat\psi_{\vec k_4}^*e^{i(\vec k_3-\vec k_4)\vec l}\\
&\times \left(e^{-i\vec k_2\vec l}+e^{i\vec k_4\vec l}-2\right)\Delta(\vec k_1-\vec k_2+\vec k_3-\vec k_4)+\text{c.c.}\Biggr].
\end{align*}

Now we use the assumption that the magnons whose wavelength is of the same order with $a$ are not essential for the considered phenomenon, in other words we assume that $ak\ll1$. In this case we can expand the exponents in exchange energy into series on $ak$, then performing the normalization we finally obtain the expressions \eqref{eq:Eex-lin} and \eqref{eq:Eex-nl} for the exchange energy.

\subsection{Dipole-dipole energy}
\label{app:Energies-Ed}

\begin{figure}
\includegraphics[width=0.85\columnwidth]{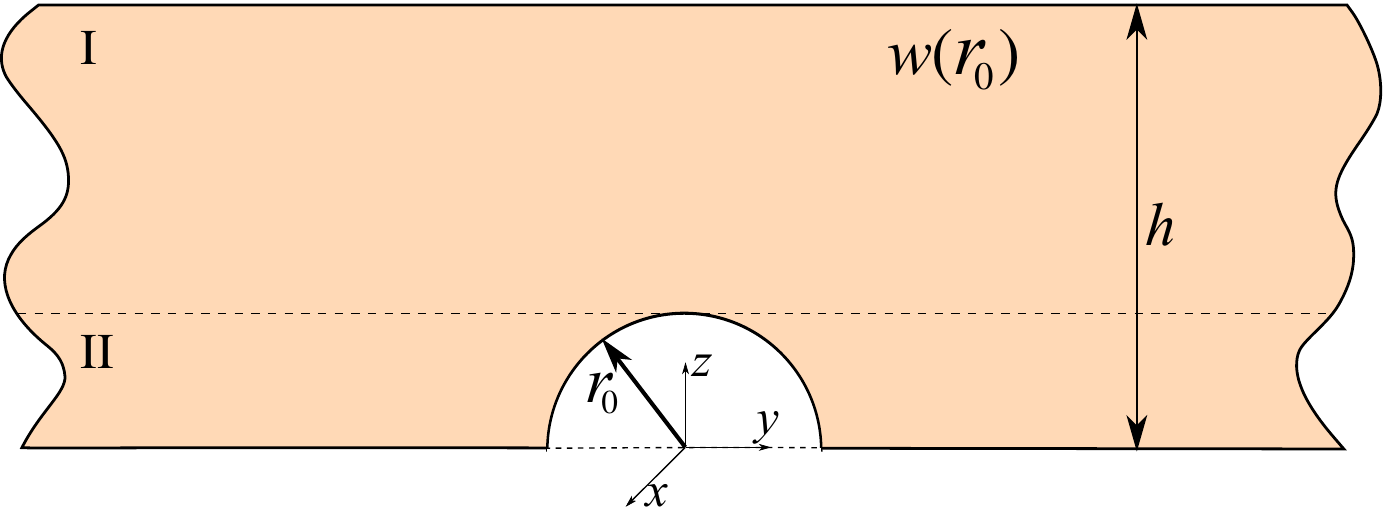}
\caption{Cross-section of the film. The filling shows domain of integration $w(r_0)$ used in the calculation of function $\hat A(\vec k)$ in \eqref{eq:hatA-lim}.}\label{fig:int-area}
\end{figure}

In case of the dipole-dipole energy we start from the general expression \eqref{eq:Ems}. Taking into account that the magnetization is uniform along $z$-coordinate one can write the energy \eqref{eq:Ems} in form\cite{Caputo07b}
\begin{equation*}
\begin{split}
E_\mathrm{d} &=-\frac{M_s^2a^6}{2}\sum\limits_{\vec n,\vec l}\bigg[A_{\vec n\vec l}\left(\vec m_{\vec n}\vec m_{\vec l}-3m_{\vec n}^zm_{\vec l}^z\right)\\
&+B_{\vec n\vec l}\left( m_{\vec n}^x m_{\vec l}^x-m_{\vec n}^ym_{\vec l}^y\right)+C_{\vec n\vec l}\left( m_{\vec n}^x m_{\vec l}^y+m_{\vec n}^ym_{\vec l}^x\right)\bigg],
\end{split}
\end{equation*}
where the coefficients $A$, $B$, and $C$ are the following
\begin{align*}
&A_{\vec n\vec l}=\frac12\sum\limits_{\begin{smallmatrix}\nu_z,\lambda_z\\ \vec\nu\ne\vec\lambda\end{smallmatrix}}\frac{(\lambda_x-\nu_x)^2+(\lambda_y-\nu_y)^2-2(\lambda_z-\nu_z)^2}{|\vec\lambda-\vec\nu|^5}\\
&B_{\vec n\vec l}=\frac32\sum\limits_{\begin{smallmatrix}\nu_z,\lambda_z\\ \vec\nu\ne\vec\lambda\end{smallmatrix}}\frac{(\lambda_x-\nu_x)^2-(\lambda_y-\nu_y)^2}{|\vec\lambda-\vec\nu|^5}\\
&C_{\vec n\vec l}=3\sum\limits_{\begin{smallmatrix}\nu_z,\lambda_z\\ \vec\nu\ne\vec\lambda\end{smallmatrix}}\frac{(\lambda_x-\nu_x)(\lambda_y-\nu_y)}{|\vec\lambda-\vec\nu|^5}.
\end{align*}
Here $\vec\nu=(\nu_x,\nu_y,\nu_z)$ and $\vec\lambda=(\lambda_x,\lambda_y,\lambda_z)$ are three-dimensional indexes while $\vec n=(\nu_x,\nu_y)$ and $\vec l=(\lambda_x,\lambda_y)$ are the corresponding two-dimensional ones. Substituting \eqref{eq:mxmymz} into the dipolar energy and taking into account that $A_{\vec n\vec l}=A_{\vec l\vec n}$, $B_{\vec n\vec l}=B_{\vec l\vec n}$ and $C_{\vec n\vec l}=C_{\vec l\vec n}$ one obtains the following expression for dipole-dipole energy $E_\mathrm{d}=E_\mathrm{d}^0+E_\mathrm{d}^\mathrm{nl}$, where
\begin{subequations} \label{eq:Ed-ABC-psi}
\begin{align}
E_\mathrm{d}^0=-\frac{M_s^2a^6}{2}\sum\limits_{\vec n,\vec l}\Biggl[&A_{\vec n\vec l}\left(2|\psi_{\vec n}|^2+\psi_{\vec n}\psi_{\vec l}^*\right)\\ \nonumber
&+ D_{\vec n\vec l}\psi_{\vec n}\psi_{\vec l}+\text{c.c.}\Biggr]\\
E_\mathrm{d}^\mathrm{nl}=\frac{M_s^2a^6}{2}\sum\limits_{\vec n,\vec l}\Biggl[&A_{\vec n\vec l}|\psi_{\vec n}|^2\left(|\psi_{\vec l}|^2+\frac12\psi_{\vec n}\psi_{\vec l}^*\right)\\ \nonumber
&+ \frac12D_{\vec n\vec l}|\psi_{\vec n}|^2\psi_{\vec n}\psi_{\vec l}+\text{c.c.}\Biggr].
\end{align}
\end{subequations}
Here the notation $D_{\vec n\vec l}=B_{\vec n\vec l}-iC_{\vec n\vec l}$ was introduced. Now we substitute \eqref{eq:four-inv} into \eqref{eq:Ed-ABC-psi} and perform the summation over $\vec n$ with taking into account \eqref{eq:orth-cond} and finally we obtain
\begin{subequations} \label{eq:Ed-ABC-psi-rec}
\begin{align}
E_\mathrm{d}^0=&-\frac{M_s^2a^6}{2}\sum\limits_{\vec k}|\Biggl\{\hat\psi_{\vec k}|^2\left[2\hat A(0)+\hat A(\vec k)\right]\\ \nonumber
&+ \hat D(\vec k)\hat\psi_{\vec k}\hat\psi_{-\vec k} + \text{c.c.}\Biggr\},\\
E_\mathrm{d}^\mathrm{nl}=&\frac{M_s^2a^6}{4\mathcal{N}_{xy}}\!\!\sum\limits_{\vec k_1\vec k_2\vec k_3\vec k_4}\!\!
\Biggl\{\left[2\hat A(\vec k_1-\vec k_2)+\hat A(\vec k_1)\right]\\ \nonumber
&\times \hat\psi_{\vec k_1}\hat\psi_{\vec k_2}^*\hat\psi_{\vec k_3}\hat\psi_{\vec k_4}^*\Delta(\vec k_1-\vec k_2+\vec k_3-\vec k_4)\\ \nonumber
&+\hat D(\vec k_1)\hat\psi_{\vec k_1}\hat\psi_{\vec k_2}^*\hat\psi_{\vec k_3}\hat\psi_{\vec k_4}\Delta(\vec k_1-\vec k_2+\vec k_3+\vec k_4) + \text{c.c.}\Biggr\}.
\end{align}
\end{subequations}
where functions $\hat A(\vec k)$ and $\hat D(\vec k)$ are determined as following
\begin{subequations} \label{eq:hatA-hatD}
\begin{align}
\label{eq:hatA}&\hat A(\vec k)=\frac12\sum\limits_{\vec l}\sum\limits_{\nu_z,\lambda_z}\frac{x_{\vec l}^2+y_{\vec l}^2-2(z_{\nu_z}-z_{\lambda_z})^2}{\left[x_{\vec l}^2+y_{\vec l}^2+(z_{\nu_z}-z_{\lambda_z})^2\right]^{5/2}}e^{i\vec l\vec k},\\
\label{eq:hatD}&\hat D(\vec k)=\frac32\sum\limits_{\vec l}\sum\limits_{\nu_z,\lambda_z}\frac{x_{\vec l}^2-y_{\vec l}^2-2ix_{\vec l}y_{\vec l}}{\left[x_{\vec l}^2+y_{\vec l}^2+(z_{\nu_z}-z_{\lambda_z})^2\right]^{5/2}}e^{i\vec l\vec k}.
\end{align}
\end{subequations}
Form of the functions $\hat A(\vec k)$ and $\hat D(\vec k)$ is not convenient for the analysis therefore we perform an approximate transition from summation to integration in \eqref{eq:hatA-hatD}. Let us start from the function $\hat A(\vec k)$:
\begin{equation} \label{eq:hatA-lim}
\hat A(\vec k)=\frac{1}{a^4}\lim\limits_{r_0\rightarrow0}\!\!\int\limits_{w(r_0)}\!\!\!\!\mathrm{d}^3\vec r(h-z)\frac{x^2+y^2-2z^2}{[x^2+y^2+z^2]^{5/2}}e^{i(xk^x+yk^y)},
\end{equation}
where we pricked out the coordinate origin from the domain of integration $w(r_0)$, see the Fig.~\ref{fig:int-area} and we also used the relation
\begin{equation}
\int\limits_0^h\mathrm{d}z\int\limits_0^h\mathrm{d}z'F(|z-z'|)=2\int\limits_0^h (h-z)F(z)\mathrm{d}z.
\end{equation}

Separating the region of integration $\omega(r_0)$ into parts I and II (see Fig.~\ref{fig:int-area}) and performing the change of variables $(x,\,y)=\rho(\cos\chi,\,\sin\chi)$ we can represent the function $\hat A$ as a sum $\hat A(\vec k)=(A_I+A_{II})/a^4$, where
\begin{subequations}\label{eq:hatA-lim-I-II}
\begin{align}
\label{eq:hatA-lim-I}&\frac{A_I}{2\pi}=\lim\limits_{\rho_0\rightarrow0}\int\limits_{\rho_0}^h\mathrm{d}z(h-z)\int_0^\infty\mathrm{d}\rho\rho\frac{\rho^2-2z^2}{[\rho^2+z^2]^{5/2}}J_0(\rho k),\\
\label{eq:hatA-lim-II}& \frac{A_{II}}{2\pi}=\lim\limits_{\rho_0\rightarrow0}\int\limits_{0}^{\rho_0}\mathrm{d}z(h-z)\!\!\!\!\!\int\limits_{\sqrt{\rho_0^2-z^2}}^\infty\!\!\!\!\!\mathrm{d}\rho\rho\frac{\rho^2-2z^2}{[\rho^2+z^2]^{5/2}}J_0(\rho k)
\end{align}
\end{subequations}
Here we performed integration over $\chi\in[0,\,2\pi]$ using the relation $\int_0^{2\pi}e^{i\rho(k^x\cos\chi+k^y\sin\chi)}\mathrm{d}\chi=2\pi J_0(k\rho)$, where $J_0(x)$ denotes zero-order Bessel function of the first kind. Direct integration in \eqref{eq:hatA-lim-I} with the consequent limit calculation results $A_I/2\pi=-hg(kh)$, where $g(x)=(e^{x}+x-1)/x$. Change of variables $\rho\rightarrow\rho_0\rho$ and $z\rightarrow\rho_0z$ allows us to get rid of the $\rho_0$ in the integration limits:
\begin{equation}
\label{eq:hatA-lim-II-chng}\frac{A_{II}}{2\pi}=\lim\limits_{\rho_0\rightarrow0}\int\limits_{0}^{1}\mathrm{d}z(h-z\rho_0)\!\!\!\!\!\int\limits_{\sqrt{1-z^2}}^\infty\!\!\!\!\!\mathrm{d}\rho\rho\frac{\rho^2-2z^2}{[\rho^2+z^2]^{5/2}}J_0(\rho\rho_0 k)
\end{equation}
Calculation the limit in \eqref{eq:hatA-lim-II-chng} with the subsequent integration results $A_{II}/2\pi=h2/3$, so finally
\begin{equation} \label{eq:hatA-final}
\hat A(\vec k)=\frac{2\pi h}{a^4}\left[\frac23-g(kh)\right].
\end{equation}

Performing the same transition to the polar coordinates $(\rho,\,\chi)$ we represent \eqref{eq:hatD} as following
\begin{equation} \label{eq:hatD-pre}
\hat D(\vec k)=\frac{1}{a^4}\int\limits_0^h\mathrm{d}z(h-z)\int\limits_0^\infty\frac{\rho^3\mathrm{d}\rho}{[\rho^2+z^2]^{5/2}}\int\limits_0^{2\pi}\mathrm{d}\chi e^{-i(\vec\rho\vec k+2\chi)},
\end{equation}
where $\vec\rho\vec k=\rho(k^x\cos\chi+k^y\sin\chi)$. The direct integration \eqref{eq:hatD-pre} using the relation $\int_0^{2\pi}\mathrm{d}\chi e^{i(x\cos\chi+n\chi)}=2\pi i^nJ_n(x)$ results
\begin{equation} \label{eq:hatD-final}
\hat D(\vec k)=-\frac{2\pi h}{a^4}g(kh)\frac{(k^x-ik^y)^2}{k^2}.
\end{equation}
Substituting now \eqref{eq:hatA-final} and \eqref{eq:hatD-final} into \eqref{eq:Ed-ABC-psi-rec} we obtain the expressions \eqref{eq:Ed-lin} and \eqref{eq:Ed-nl} for the dipole-dipole energy.

\section{Stability of the vortex-antivortex lattice solution}\label{app:stability}
Here we consider stability of the stationary solution~\eqref{eq:N-Phi-cond}, \eqref{eq:N-cosPhi-final} of the system \eqref{eq:eq-motion-N-Phi}. As it was sown in the main text after period of time $\tau=1/(2\varkappa)$ the solutions of Eqs.~\eqref{eq:eq-motion-N-Phi} satisfy the conditions $N_\uparrow=N_\downarrow$ and $N_\rightarrow=N_\leftarrow$. Using these conditions and introducing variables $N_1=N_\uparrow=N_\downarrow$, $N_2=N_\rightarrow=N_\leftarrow$, $\Phi_1=\Phi_\uparrow+\Phi_\downarrow$ and $\Phi_2=\Phi_\rightarrow+\Phi_\leftarrow$ one can reduce the system of eight equations \eqref{eq:eq-motion-N-Phi} to the system of four equations
\begin{equation}\label{eq:system-four}
\begin{split}
&\dot{N}_i=-\frac{\partial\mathcal{E}}{\partial\Phi_i}-F^N_i,\\
&\dot{\Phi}_i=\frac{\partial\mathcal{E}}{\partial N_i}-F^\Phi_i,\qquad i=1,\,2.
\end{split}
\end{equation}
Here the forces of the spin-current acting has the form
\begin{equation}
\begin{split}
F^N_i=&2\varkappa \left\{N_i\left[1-\frac{3N_i+4N_{\bar i}}{2\Lambda^2\mathcal{N}_{xy}}\right]-\frac{N_iN_{\bar i}}{\Lambda^2\mathcal{N}_{xy}}\cos(\Phi_i-\Phi_{\bar i})\right\},\\
F^\Phi_i=&\frac{2\varkappa}{\Lambda^2\mathcal{N}_{xy}}N_{\bar i}\sin(\Phi_i-\Phi_{\bar i}),
\end{split}
\end{equation}
where the notation $\bar1=2$ and $\bar2=1$ is used. Hamiltonian \eqref{eq:Hamiltonian-N-Phi} in terms of the new variables takes the form
\begin{subequations}
\begin{align}
\mathcal{E}=\mathcal{E}^0+\mathcal{E}^\mathrm{nl},
\end{align}
where the linear part \eqref{eq:Hamiltonian-N-Phi-lin} reads
\begin{align}
\mathcal{E}^0&=2(N_1+N_2)\left(K^2\ell^2+\frac{g_1}{2}-1\right)-\\ \nonumber
&-g_1\left(N_1\cos\Phi_1-N_2\cos\Phi_2\right)
\end{align}
and the corresponding nonlinear part \eqref{eq:Hamiltonian-N-Phi-nl} is
\begin{align}
&\mathcal{E}^\mathrm{nl}=\frac{1}{\mathcal{N}_{xy}}\Biggl\{\sum\limits_{i=1,2}\Bigl[N_i^2(3+K^2\ell^2-\frac32g_1-g_2)+\\ \nonumber
&+2N_iN_{\bar i}\Bigl(\cos(\Phi_i-\Phi_{\bar i})(1+K^2\ell^2-\frac{g_1}{2}-g_{\sqrt2})+ \\ \nonumber
&+2-g_1-g_{\sqrt2}\Bigr)\Bigr]+ \frac32g_1\left(N_1^2\cos\Phi_1-N_2^2\cos\Phi_2\right)+\\ \nonumber
&g_1N_1N_2(\cos\Phi_1-\cos\Phi_2)\Biggr\}.
\end{align}
\end{subequations}
Now we linearize the system \eqref{eq:system-four} against a stationary solution $\vec{\mathrm{v}}^0=\{N_1^0,\,N_2^0,\,\Phi_1^0,\,\Phi_2^0\}$:
\begin{equation}
\dot{\tilde{\vec{\mathrm{v}}}}=\mathbf{M}\tilde{\vec{\mathrm{v}}},
\end{equation}
where $\tilde{\vec{\mathrm{v}}}$ is small deviation from the solution $\vec{\mathrm{v}}^0$ and  $4\times4$ matrix $\mathbf{M}$ can be presented in the following block form
\begin{subequations}
\begin{align}
\mathbf{M}=\begin{pmatrix}
\mathbf{M}^{NN}&\mathbf{M}^{N\Phi}\\
\mathbf{M}^{\Phi N}&\mathbf{M}^{\Phi\Phi}
\end{pmatrix}_{\vec{\mathrm{v}}=\vec{\mathrm{v}}^0},
\end{align}
where the components are the following $2\times2$ matrixes
\begin{align}\label{eq:matrix-components}
\mathbf{M}^{NN}_{i,j}&=-\frac{\partial^2\mathcal{E}}{\partial N_i\partial N_j}-\frac{\partial F^N_i}{\partial N_j},\\ \nonumber
\mathbf{M}^{N\Phi}_{i,j}&=-\frac{\partial^2\mathcal{E}}{\partial N_i\partial \Phi_j}-\frac{\partial F^N_i}{\partial \Phi_j},\\ \nonumber
\mathbf{M}^{\Phi N}_{i,j}&=\frac{\partial^2\mathcal{E}}{\partial \Phi_i\partial N_j}-\frac{\partial F^\Phi_i}{\partial N_j},\\ \nonumber
\mathbf{M}^{\Phi\Phi}_{i,j}&=\frac{\partial^2\mathcal{E}}{\partial \Phi_i\partial \Phi_j}-\frac{\partial F^\Phi_i}{\partial \Phi_j},\qquad i,j=1,2.
\end{align}
\end{subequations}
Necessary and sufficient condition for stability of the solution $\vec{\mathrm{v}}^0$ is negativity of real parts of all eigenvalues $\lambda_1,\,\lambda_2,\,\lambda_3,\,\lambda_4$ of the matrix $\mathbf{M}$. After the straightforward calculation of \eqref{eq:matrix-components} we substitute the solution $N_1=N_2=N$, $\Phi_1=\Phi$ and $\Phi_2=\Phi-\pi$ what corresponds to the conditions \eqref{eq:N-Phi-cond}. Then after the straightforward calculation of the eigenvalues of $\mathbf{M}$ we exclude $\Phi$ using \eqref{eq:N-Phi-final} and  finally we obtain
\begin{equation}\label{eq:eigenvalues}
\begin{split}
&\lambda_1=-\varkappa-\sqrt{2\varkappa_c^2-\varkappa^2}+\frac{2\mathscr{N}}{\varkappa_c}\left[\mathfrak{FG}+\varkappa_c^2\left(3+\frac{1}{\Lambda^2}\right)\right],\\
&\lambda_2=-\varkappa+\sqrt{2\varkappa_c^2-\varkappa^2}-\frac{2\mathscr{N}}{\varkappa_c}\left[\mathfrak{FG}+3\varkappa_c^2\left(1-\frac{1}{\Lambda^2}\right)\right],\\
&\lambda_3=-\varkappa+\sqrt{2\varkappa_c^2-\varkappa^2}-\frac{2\mathscr{N}}{\varkappa_c}\left[\mathfrak{FG}+5\varkappa_c^2\left(1-\frac{1}{\Lambda^2}\right)\right],\\
&\lambda_4=-\varkappa-\sqrt{2\varkappa_c^2-\varkappa^2}+\frac{2\mathscr{N}}{\varkappa_c}\left[\mathfrak{FG}+5\varkappa_c^2\right],
\end{split}
\end{equation}
where only the terms linear with respect to $\mathscr{N}$ are saved. It should be noted that for $\mathscr{N}=0$ (what corresponds to the saturated state) the condition $\lambda_i<0$ in \eqref{eq:eigenvalues} is equivalent to the condition of stability of the saturated state $\varkappa>\varkappa_c$. Substituting \eqref{eq:N-final} into \eqref{eq:eigenvalues} and considering the indefinitely small deviation from the saturation current $\varkappa=\varkappa_c+\delta$ one obtains the following linear approximation of \eqref{eq:eigenvalues} with respect to deviation $\delta$
\begin{equation}\label{eq:eigen-dev}
\begin{split}
&\lambda_1=-2\left[\varkappa_c+2\delta\frac{\mathfrak{FG}+\varkappa_c^2\left(3+\frac{1}{\Lambda^2}\right)}{\mathfrak{FG}+5\varkappa_c^2\left(1-\frac{1}{\Lambda^2}\right)}\right],\\
&\lambda_2=4\delta\left[\frac{\mathfrak{FG}+3\varkappa_c^2\left(1-\frac{1}{\Lambda^2}\right)}{\mathfrak{FG}+5\varkappa_c^2\left(1-\frac{1}{\Lambda^2}\right)}-\frac12\right],\\
&\lambda_3=2\delta,\\
&\lambda_4=-2\left[\varkappa_c+2\delta\frac{\mathfrak{FG}+5\varkappa_c^2}{\mathfrak{FG}+5\varkappa_c^2\left(1-\frac{1}{\Lambda^2}\right)}\right].
\end{split}
\end{equation}
Taking into account that $\mathfrak{FG}>0$ and $\Lambda>1$ one can conclude that all $\lambda_i<0$ only for $\delta<0$. That means that the obtained lattice solution  \eqref{eq:N-Phi-cond}, \eqref{eq:N-cosPhi-final} is stable for infinitely small decrease in current from the critical value.
%

\end{document}